# PARAMETRIC-RESONANCE IONIZATION COOLING OF MUON BEAMS

Ya.S. Derbenev, V.S. Morozov
*Jefferson Lab, Newport News, VA 23606, USA*
A. Afanasev
*The George Washington University, Washington, DC 20052, USA*
K.B. Beard, R.P. Johnson
*Muons, Inc., Batavia, IL 60610, USA*
B. Erdelyi, J.A. Maloney
*Northern Illinois University, DeKalb, IL 60115, USA*

*Abstract*

Cooling of muon beams for the next-generation lepton collider is necessary to achieve its higher luminosity with fewer muons. In this paper we present an idea to combine ionization cooling with parametric resonances that is expected to lead to muon beams with much smaller transverse sizes. We describe a linear magnetic transport channel where a half integer resonance is induced such that the normal elliptical motion of particles in $x$-$x'$ phase space becomes hyperbolic, with particles moving to smaller $x$ and larger $x'$ at the channel focal points. Thin absorbers placed at the focal points of the channel then cool the angular divergence of the beam by the usual ionization cooling mechanism where each absorber is followed by RF cavities. We present a theory of Parametric-resonance Ionization Cooling (PIC), starting with the basic principles in the context of a simple quadrupole-focused beam line. Then we discuss detuning caused by chromatic, spherical, and non-linear field aberrations and the techniques needed to reduce the detuning. We discuss the requirement that PIC be accompanied by emittance exchange in order to keep the momentum spread sufficiently small. Examples of PIC channel are presented, along with computer simulations aimed at practical implementation of the described theoretical concept.

# I. INTRODUCTION

Experiments at energy-frontier colliders require high luminosities, of order $10^{34}$ cm$^{-2}$ sec$^{-1}$ or more, in order to obtain reasonable rates for events having point-like cross sections. High luminosities require intense beams, small transverse emittances, and a small beta function at the collision point. For muon colliders, high beam intensities and small emittances are difficult and expensive to achieve because muons are produced diffusely and must be cooled drastically within their short lifetimes. Ionization cooling as it is presently envisioned will not cool the beam sizes sufficiently well to provide adequate luminosity without large muon intensities. However, to the extent that the transverse emittances can be reduced further than with conventional ionization cooling, several problems can be alleviated.

Lower transverse emittance allows a reduced muon current for a given luminosity, which implies:
1) a proton driver with reduced demands to produce enough proton power to create the muons;
2) reduced site boundary radiation limits from circulating muons that decay into neutrinos that interact in the earth;
3) reduced detector background from the electrons from the decay of circulating muons;
4) reduced proton target heat deposition and radiation levels;
5) reduced heating of the ionization cooling energy absorber;
6) less beam loading and wake field effects in the accelerating RF cavities.

Smaller transverse emittance has virtues beyond reducing the required beam currents, namely:
1) smaller, higher-frequency RF cavities with higher gradient can be used for acceleration;
2) beam transport is easier, with smaller aperture magnetic and vacuum systems;
3) stronger collider interaction point focusing can be used, since that is limited by beam extension in the IP quadrupoles.

Ionization cooling of a muon beam involves passing a magnetically focused beam through an energy absorber, where the muon transverse and longitudinal momentum components are reduced, and through RF cavities, where only the longitudinal component is regenerated. After some distance, the transverse components shrink to the point where they come into equilibrium with the heating caused by multiple coulomb scattering. The equation describing the rate of cooling is a balance between these cooling (first term) and heating (second term) effects:

$$\frac{d\varepsilon_n}{ds} = -\frac{1}{v^2}\frac{dE_\mu}{ds}\frac{\varepsilon_n}{E_\mu} + \frac{1}{v^3}\frac{\beta(0.014)^2}{2E_\mu m_\mu X_0}.$$

Here $\varepsilon_n$ is the normalized emittance, $E_\mu$ is the muon energy in GeV, $dE_\mu/ds$ and $X_0$ are the energy loss and radiation length of the absorber medium, $\beta = \lambda/2\pi$ is the transverse beta-function of the magnetic channel, and $v$ is the particle velocity normalized to light velocity.

Setting the heating and cooling terms equal defines the equilibrium emittance, the very smallest possible with the given parameters:

$$\varepsilon_n^{(equ.)} = \frac{\beta(0.014)^2}{2vm_\mu \frac{dE_\mu}{ds} X_0}.$$

One can see that the figure of merit for a cooling absorber material is the product of the energy loss rate times the scattering length. Up to now, liquid hydrogen has been the energy-absorbing medium of choice, with $dE/ds$= 60 MeV/m and $X_0$ = 8.7 m. Superconducting solenoidal focusing is used to give a small value of $\beta$ : 10 cm, corresponding to a 100 MeV/c muon in a 6 T field.

In the PIC technique described below, the resonant approach to particle focusing can achieve equilibrium transverse emittances that are at least an order of magnitude smaller than in conventional ionization cooling. That is, the beam cooling in a conventional ionization cooling channel would require an unrealistically high magnetic field for an equilibrium transverse emittance as small as PIC can achieve. Another advantage to PIC is that a beryllium energy absorber is more effective than hydrogen, since the ratio of the absorber thickness to the betatron wavelength is a measure of its effectiveness. That is, the density of beryllium relative to hydrogen is more important than the product of energy loss rate times Coulomb scattering length in a channel that is required to be as short as possible to reduce losses by muon decay. The practical advantage of beryllium is that the required energy absorber geometry and refrigeration have straightforward engineering solutions.

## II. BASIC PIC CONCEPTS

**A. Hyperbolic dynamics of PIC**

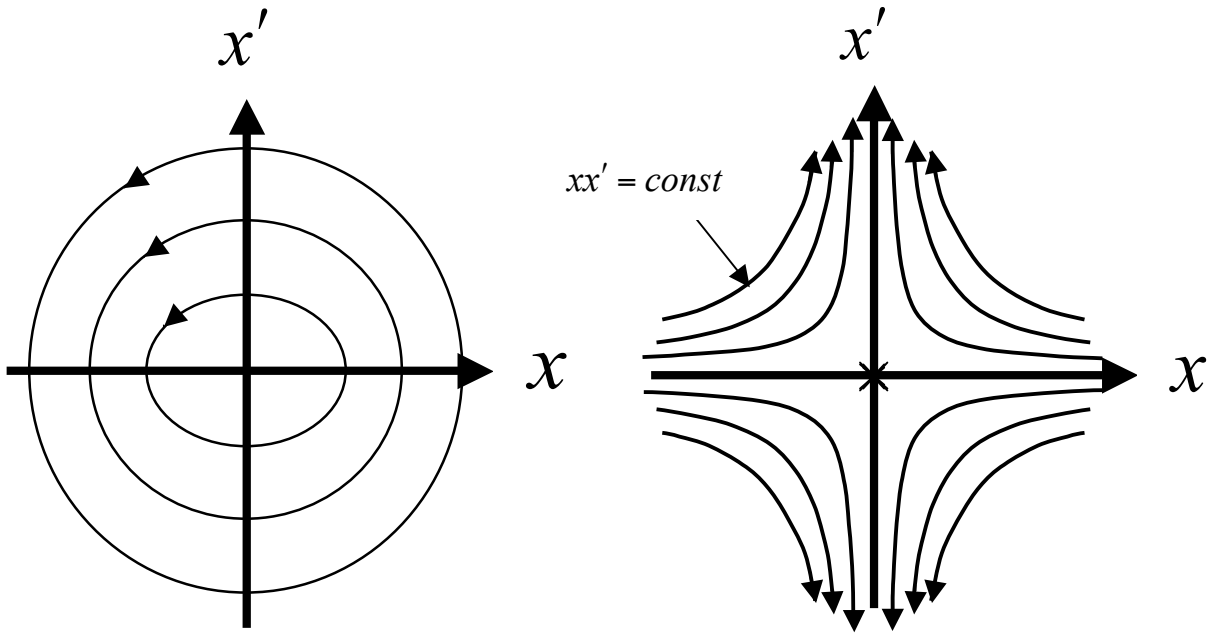

**Figure 1:** Comparison of particle motion at periodic locations along the beam trajectory in transverse phase space for: LEFT ordinary oscillations and RIGHT hyperbolic motion induced by perturbations at a harmonic of the betatron frequency.

In general, a parametric resonance is induced in an oscillating system by using a perturbing frequency that is the same as or a harmonic of a parameter of the system. Physicists are often first introduced to this phenomenon in the study of a rigid pendulum, where a periodic perturbation of the pivot point can lead to stable motion with the pendulum upside down. Half-integer resonant extraction from a synchrotron is another example familiar to accelerator physicists, where larger and larger radial excursions of particle orbits at successive turns are induced by properly placed quadrupole magnets that perturb the beam at a harmonic of the betatron frequency. In this case, the normal elliptical motion of a particle's horizontal coordinate in phase space at the extraction septum position becomes hyperbolic, $xx' = const$, leading to a beam emittance which has a wide spread in $x$ and narrow spread in $x'$.

In PIC, the same principle is used but the perturbation generates hyperbolic motion such that the emittance becomes narrow in $x$ and wide in $x'$ at certain positions as the beam passes down a line or circulates in a ring. Ionization cooling is then used to damp the angular spread of the beam. Figure 1 shows how the motion is altered by the perturbation. The principle of ionization cooling [1] is well known, where a particle loses momentum in all three coordinates as it passes through some energy absorbing material and only the longitudinal component is replaced by RF fields. The angular divergence $x'$ of the particle is thereby reduced until it reaches equilibrium with multiple Coulomb scattering in the material. The concept of ionization cooling is shown in Figure 2.

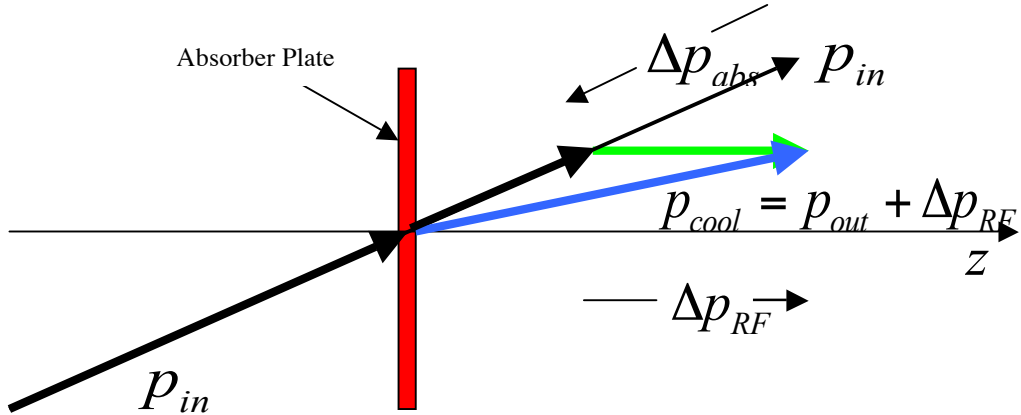

**Figure 2:** Principle of transverse ionization cooling. A particle loses momentum in all three coordinates as it passes through an energy absorbing plate. Only the longitudinal component is replaced by RF fields, thereby reducing the angular divergence of the particle, $\vartheta = x'$.

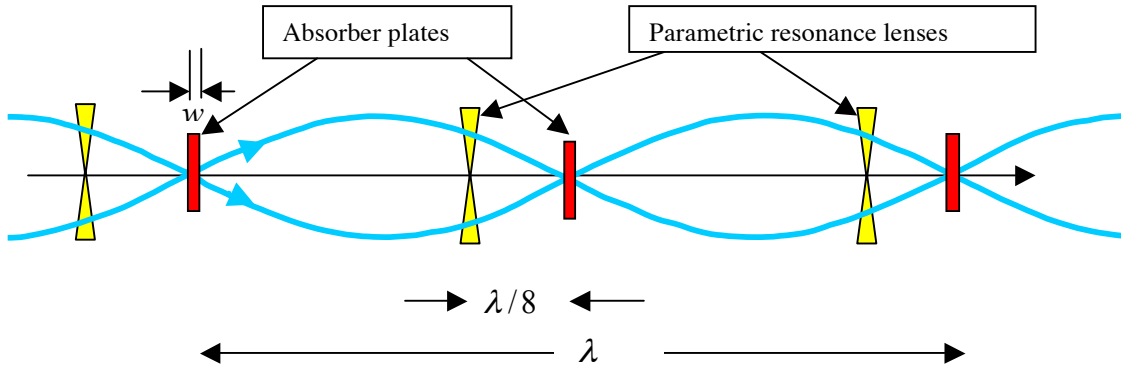

**Figure 3:** Conceptual diagram of a beam cooling channel in which hyperbolic trajectories are generated in transverse phase space by perturbing the beam at the betatron frequency, a parameter of the beam oscillatory behavior. Neither the focusing magnets that generate the betatron oscillations nor the RF cavities that replace the energy lost in the absorbers are shown in the diagram. The blue trajectories indicate the betatron motion of particles that define the beam envelope.

Thus in PIC the phase space area is reduced in $x$ due to the dynamics of the parametric resonance and $x'$ is reduced or constrained by ionization cooling. For PIC to work, however, the beam must be cooled first by other means. For this analysis the initial conditions for PIC are assumed to be those as might be attained using a helical cooling channel [2] as shown in the table below.

| Parameter | Unit | equilibrium rms value |
| --- | --- | --- |
| Beam momentum, $p$ | MeV/c | 100 |
| Synchrotron emittance, $\varepsilon_s$ | $\mu m$ | 300 |
| Relative momentum spread | % | 2 |
| Beam width due to $\Delta p/p$ | mm | 1.5 |
| Bunch length | mm | 11 |
| Transverse emittances, $\varepsilon_+/\varepsilon_-$ | mm-mr | 100/300 |
| Beam widths, $\sigma_1/\sigma_2$ | mm | 4.5/2.8 |

**Table1:** Beam parameters at the output of a proposed 6D helical cooling channel [2].

Let there be a periodic focusing lattice of period $\lambda$ along the beam path with coordinate $z$. Particle tracking or mapping is based on a single period transformation matrix, M (between two selected points, $z_0$ and $z_0 + \lambda$), for particle transverse coordinate and angle,

$$\begin{pmatrix} x \\ x' \end{pmatrix}_{z_0+\lambda} = M_x \begin{pmatrix} x \\ x' \end{pmatrix}_{z_0},$$

with a similar expression for the y coordinate.

The matrices $M_x$ and $M_y$ are symplectic or canonical, which means each has determinant equal to one. Otherwise, the matrix elements are arbitrary in general. Thus, each can be represented in a general form convenient for later discussions as follows:

$$M = \begin{pmatrix} e^{-\lambda\Lambda_d}\cos\psi & g\sin\psi \\ -\frac{1}{g}\sin\psi & e^{\lambda\Lambda_d}\cos\psi \end{pmatrix} \quad (1)$$

In particular, the optical period can be designed such that, for $\sin\psi = 0$ (i.e. $\psi = \pi$ or $\psi = 2\pi$), the evolving particle coordinate and angle (or momentum) appear uncoupled:

$$(x)_{z_0+\lambda} = \pm e^{-\lambda\Lambda_d}(x)_{z_0} \text{ and } (x')_{z_0+\lambda} = \pm e^{\lambda\Lambda_d}(x')_{z_0}.$$

Thus, if the particle angle at point $z_0$ grows ($\Lambda_d > 0$), then the transverse position experiences damping, and vice versa. Liouville's theorem is not violated, but particle trajectories in phase space are hyperbolic ($xx' = const$); this is an example of a parametric resonance. Exactly between the two resonance focal points the opposite situation occurs where the transverse particle position grows from period to period, while the angle damps.

### B. Stabilizing absorber effect

If we now introduce an energy absorber plate of thickness $w$ at each of the resonance focal points as shown in Figure 3, ionization cooling damps the angle spread with a rate $\Lambda_c$. Here we assume balanced 6D ionization cooling, where the three partial cooling decrements have been equalized using emittance exchange techniques as described in reference [2]:

$$\Lambda_c = \frac{1}{3}\Lambda, \quad \Lambda = 2\frac{\langle \gamma'_{abs} \rangle}{\gamma} = 2\frac{\gamma'_{acc}}{\gamma}, \quad \langle \gamma'_{abs} \rangle = \gamma'_{abs} 2w/\lambda,$$

where $\gamma'_{abs}$ and $\gamma'_{acc}$ are the intrinsic absorber energy loss and the RF acceleration rates, respectively. If $\Lambda_d = \Lambda_c/2$ then the angle spread and beam size are damped with decrement $\Lambda_c/2$:

$$\begin{pmatrix} x \\ x' \end{pmatrix}_{z_0+\lambda} = e^{-\lambda\Lambda_c/2} \begin{pmatrix} x \\ x' \end{pmatrix}_{z_0}.$$

### C. Reduction of phase diffusion and equilibrium emittance

The rms angular spread is increased by scattering and decreased by cooling,

$$\frac{d}{dz}\overline{(x')^2} = \frac{(Z+1)}{2\gamma v^2}\frac{m_e}{m_\mu}\Lambda - \Lambda_c\overline{(x')^2},$$

which leads to the equilibrium angular spread at the focal point:

$$\overline{(x')^2}_{eq} = \frac{1}{\Lambda_c}\frac{d}{dz}\overline{(x')^2}_{z_0} = \frac{3}{2}\frac{(Z+1)}{\gamma v^2}\frac{m_e}{m_\mu}. \tag{2}$$

The rms product $\left[\overline{(x^2)}\cdot\overline{(x')^2}\right]^{\frac{1}{2}}_{z_0}$ determines the effective 2D beam phase space volume, or emittance.

Taking into account the continuity of x in collisions, the diffusion rate of particle position at the focus is a function of $s = z - z_0$, the local position of the beam within the absorber:

$$\delta(x)_{z_0} = -s\delta x', \quad -\frac{w}{2} \leq s \leq \frac{w}{2},$$

$$\frac{d}{dz}\overline{(\delta x)^2}_{z_0} = \frac{w^2}{12}\frac{d}{dz}\overline{(\delta x')^2}. \tag{3}$$

Thus, in our cooling channel with resonance optics and correlated absorber plates, the equilibrium beam size at the plates is determined not by the characteristic focal parameter of the optics, $\lambda/2\pi$, but by the thickness of absorber plates, $w$. Hence, the equilibrium emittance is equal to

$$(\varepsilon_x)_{eq} = \gamma v \left[\overline{(x^2)}\cdot\overline{(x')^2}\right]^{\frac{1}{2}}_{z_0} = \gamma v \frac{w}{2\sqrt{3}}\overline{(x')^2}_{z_0} = \frac{\sqrt{3}}{4v}(Z+1)\frac{m_e}{m_\mu}w.$$

The emittance reduction by PIC is improved compared to a conventional cooling channel by a factor

$$\frac{\pi}{\sqrt{3}}\frac{w}{\lambda} = \frac{\pi}{2\sqrt{3}}\frac{\gamma'_{acc}}{\gamma'_{abs}}.$$

Using the well-known formula for the instantaneous energy loss rate in an absorber, we find an explicit expression for the transverse equilibrium emittance that can be achieved using PIC:

$$\varepsilon_x = \frac{\sqrt{3}}{16}v\left(1+\frac{1}{Z}\right)\frac{(\lambda/2\pi)}{nr_e^2 \log}\gamma'_{acc}.$$

Here $Z$ and $n$ are the absorber atomic number and concentration, $r_e$ the classical electron radius, and $v$ is the muon velocity. Here log is a symbol for the Coulomb logarithm of ionization energy loss for fast particles:

$$\log \equiv \ln\left(\frac{2p^2}{h\nu\, m_\mu}\right) - v^2,$$

with $m_\mu$ the muon mass and $h\nu$ the effective ionization potential [3]. A typical magnitude of the log is about 12 for our conditions. The equilibrium emittance in the resonance channel is primarily determined by the absorber atomic concentration, and it decreases with beam energy in the non-relativistic region.

### III. COMBINING PIC WITH EMITTANCE EXCHANGE

Longitudinal cooling must be used with PIC in order to maintain the energy spread at the level achieved by the basic 6D cooling from the helical cooling channel. Emittance exchange must therefore be used for longitudinal cooling, which requires the introduction of bends and dispersion. Since the beam has already undergone basic 6D cooling, its transverse sizes are so small that the absorber plates can have a large wedge angle to provide balanced 6D cooling even with small dispersion. Since the beam size at the absorber plates decreases as it cools, the wedge angle can increase along the beam path while the dispersion decreases. In this way, longitudinal cooling is maintained while the straggling impact on transverse emittance is negligible. Thus there is no conceptual contradiction to have simultaneous maximum transverse PIC with optimum longitudinal cooling.

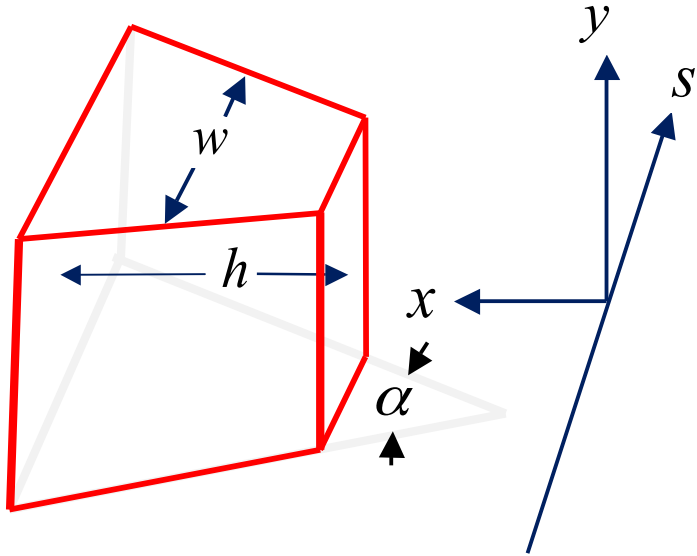

**Figure 4:** Geometry of a wedge-shaped absorber. The average thickness of the absorber in the beam ($s$) direction is $w$. In this diagram, the magnetic field of the dipoles of the channel is along the $y$ axis such that the dispersion is in the $x$ direction, where $h$ is the horizontal size of the absorber in the $x$ direction. The wedge angle $\alpha$ is determined by the requirement that emittance exchange and longitudinal cooling must accompany PIC to keep the energy spread small to control chromatic detuning.

## A. Emittance exchange using wedge absorbers

In order to prevent energy spread growth in the beam due to energy straggling in the absorber, one can use wedge absorber plates and introduce dispersion, i.e. make the beam orbit energy-dependent. Such dependence results from a beam bend by a dipole field (alternating along the beam line). As usual, the particle coordinate relative to a reference orbit can be represented as a superposition

$$x = D\frac{\Delta\gamma}{\gamma v^2} + x_b \quad (4)$$

where the dispersion $D$ and $x_b$ do not interfere on the particle trajectory. The absorber wedge orientation (i.e. the gradient of the plate width) must alternate coherently with the $D$ oscillation. Considering only the effects of energy loss in the wedge absorber plates, we find a systematic change of particle energy and position $x_b$ at the plates:

$$\Delta\gamma' = \frac{\partial\gamma'}{\partial\gamma}\Delta\gamma + \frac{\partial\gamma'}{\partial x}x = \frac{\Lambda}{2v^2}(\frac{2}{\gamma^2} - \frac{D_a}{h}\Delta\gamma)$$

$$(x_{0b}^2)' = -2(\Lambda_d - \frac{\Lambda}{2v^2}\frac{D_a}{h})x_{0b}^2.$$

Here we introduced the parameter $h$ as an effective height of the absorber wedge as indicated in figure 4:

$$h^{-1} = \frac{1}{\gamma'}\frac{\partial\gamma'}{\partial x},$$

and $D_a$, the dispersion at the absorber plates. Thus, if the ratio $D_a/h$ is positive, there is damping of the energy spread while the phase cooling decrement decreases. Let us assume an arrangement that makes the decrements of the three emittances equal to $\Lambda/3$ yet leaves equal the damping decrements of beam size and angle spread at the absorber plates. This assumption leads to the following relationships:

$$D_a = 2h(1 - \frac{2}{3}v^2) \equiv D_o. \quad (5)$$

## IV. PIC FOR AN IDEALLY TUNED BEAM

### A. Transverse equilibrium

The dispersion introduced for longitudinal cooling will also cause transverse emittance growth because of straggling, the stochastic change of particle energy due to scattering off of electrons in the absorber [4]. The related change of the particle 'free' coordinate $x_b$ after scattering in a plate can be found simply taking again into account the continuity of the total coordinate x:

$$D_0\frac{\delta\gamma}{\gamma v^2} + \delta x_b = 0.$$

Thus we find the betatron coordinate diffusion rate due to energy straggling:

$$\frac{d}{ds}\overline{(\delta x_b)_{str}^2} = \frac{1}{2}\overline{\frac{D^2}{\gamma^2 v^4}\frac{d}{ds}(\delta\gamma)^2} = \frac{\Lambda}{8}\frac{\gamma^2+1}{\gamma v^2}\frac{m_e}{\log m_\mu}D_a^2.$$

Combining this with the angle scattering in equation (3), we obtain the total rate of betatron coordinate diffusion and the associated beam size at the absorber plate and the emittance:

$$\overline{(x_b^2)'} = \frac{\Lambda}{8\gamma\beta^2} \frac{m_e}{m_\mu} \left[ \frac{Z+1}{3} w^2 + \frac{\gamma^2+1}{\log} 4h^2 (1-\frac{2}{3}v^2)^2 \right]$$

$$\sigma_b^2 = \frac{1}{8\gamma v^2} \frac{m_e}{m_\mu} [(Z+1)w^2 + 12 \frac{\gamma^2+1}{\log} h^2 (1-\frac{2}{3}v^2)^2] \qquad (6)$$

$$\varepsilon_{eq} = \varepsilon_0 \equiv \gamma v \vartheta_0 \sigma_0$$

The contribution of straggling to transverse emittance growth can be minimized by reducing the wedge absorber horizontal size $h$. However $h$ must be large compared to $\sigma_b$. Let us introduce $\chi_h = \sigma_b / h$, then equation (6) can be written:

$$\sigma_b^2 = \frac{Z+1}{8\gamma v^2} \frac{m_e}{m_\mu} w^2 \frac{1}{1 - \frac{3}{2\chi^2 v^2} \frac{m_e}{m_\mu} \frac{\gamma^2+1}{\gamma \log} (1-\frac{2}{3}v^2)^2}$$

The condition for straggling not to be important is

$$1 \gg \chi \gg (1-\frac{2}{3}v^2)[\frac{3}{2v^2} \frac{\gamma^2+1}{\gamma \log} \frac{m_e}{m_\mu}]^{1/2}$$

**B. Optimum cooling and PIC equilibrium**

We define optimum cooling by equating the three emittance cooling rates (thus, making each of them equal to $\Lambda/3$) to obtain the following relationships:

$$\frac{D}{h} = 2 - \frac{4}{3}v^2; \quad \frac{\Lambda_d}{\Lambda} = \frac{1}{2v^2} - \frac{1}{6} \qquad (7)$$

Then the balance equations lead to equilibrium as follows:

$$\theta^2 = \frac{3m_e}{2\gamma v^2 m_\mu} (Z+1+\frac{\gamma^2+1}{4\log} D'^2) \qquad (8)$$

$$\sigma^2 = \frac{3m_e}{2\gamma v^2 m_\mu} (\frac{Z+1}{12} w^2 + \frac{\gamma^2+1}{4\log} D^2) \qquad (9)$$

$$(\frac{\Delta p}{p})^2 = \frac{3m_e}{8\gamma v^2 m_\mu} \cdot \frac{\gamma^2+1}{\log}.$$

For the conditions that follow from equations (8) and (9),

$$D' \ll 2\left(\frac{Z+1}{\gamma^2+1}\log\right)^{1/2} \text{ and } h = \frac{w}{(1-\frac{2}{3}v^2)}\left[\frac{(Z+1)\log}{12(\gamma^2+1)}\right]^{1/2},$$

the equilibrium normalized transverse emittance and beam size $\sigma$ will be close to minimum values:

$$\varepsilon_\perp \Rightarrow \varepsilon_{\perp 0} = \frac{\sqrt{3}}{4v}(Z+1)\frac{m_e}{m_\mu}w, \quad \sigma = \frac{\theta w}{2\sqrt{3}} \quad (10)$$

## C. PIC potential

The equilibrium emittance (5) can be expressed as function of the intrinsic energy loss in the absorber, $E_i'$ and the average energy loss or accelerating field using the relationships $(2w/\lambda)E_i' = <E_i'> = <E_{acc}'>$:

$$\varepsilon_{\perp 0} = \frac{\sqrt{3}}{8v}\frac{m_e}{m_\mu}\lambda\frac{<E_i'>}{E_i'}(Z+1).$$

The emittance that can be achieved after an "ordinary" (non-resonance) 6D cooling [1] in hydrogen absorber is $\varepsilon_{ord} = (3\lambda m_e / 2\pi v m_\mu)$; so PIC results in the reduction of transverse emittance by a factor $\frac{\pi}{4\sqrt{3}}\frac{<E_i'>}{E_i'}(Z+1)$. As a function of $Z$, the factor $E_i'/(Z+1)$ is about $(60/v^2)$ MeV/m in case of beryllium (compare with $(15/v^2)$ MeV/m in case of liquid hydrogen) and does not change significantly with $Z$ for heavier elements. Note that use of absorbers with large atomic number is disadvantageous because of the small thickness of the plates, which makes it difficult to tune to resonance. Note also that the equilibrium emittance can be decreased by decreasing the plate thickness and lowering the accelerating field, which makes the beam line longer. Thus the cooling rate and equilibrium emittance are limited by beam loss due to muon decay. For an optimal PIC design, the plate thickness $w$ should diminish along the cooling channel, starting from a maximum determined by the available accelerating voltage. Table 2 illustrates the PIC effect.

Table 2: Potential PIC effect

| Parameter | Unit | Initial | Final |
|---|---|---|---|
| Beam momentum, p | MeV/c | 100 | 100 |
| Distance between plates, $\lambda/2$ | cm | 19 | 19 |
| Plate thickness, $w$ | mm | 6.4 | 1.6 |
| Intrinsic energy loss rate (Be) | MeV/m | 600 | 600 |
| Average energy loss | MeV/m | 20 | 5 |
| Angle spread at plates, rms, $\theta_x = \theta_y$ | mr | 150 | 200 |
| Beam transverse size at plates, rms, $\sigma_x = \sigma_y$ | mm | 2.0 | 0.1 |
| Transverse rms emittance, norm. $\varepsilon_x = \varepsilon_y$ | μm | 300 | 20 |
| Momentum spread, $\Delta p/p$, rms | % | 2.7 | 2.7 |
| Bunch length $\sigma_z$, rms | cm | 1 | 1 |
| Longitudinal emittance $\sigma_z \Delta p/mc$ | cm | 2.7x10$^{-2}$ | 2.7x10$^{-2}$ |
| PIC channel length | m | 100 ||
| Integrated energy loss | GeV | 0.7 ||
| Beam loss due to muon decay | % | 15 ||
| Number of particles/bunch |  | 10$^{11}$ ||
| Tune spread due to space charge*, $\Delta\lambda/\lambda$ | .001 | 0.75 ||

*To overcome the space charge impact on tuning, one can implement a beam recombining scheme, namely, generate a low charge/bunch beam and recombine bunches after cooling and acceleration to sufficiently relativistic energy (under investigation).

## V. TUNING DEMANDS OF PIC

For PIC to work at all, the relative spread in betatron function must be much smaller than the ratio of the betatron function to the cooling length, $l_c = \gamma mc^2/<E'>$:

$$\frac{\Delta\beta}{\beta} << \frac{\beta}{l_c}.$$

The phase advance accumulated along a single cooling decrement length must be smaller than the absorber thickness divided by the beta function in order for PIC to work with full efficiency:

$$\delta\psi = \int_{l_c} \frac{\delta\beta(s)}{\beta^2} ds << \frac{w}{\beta},$$

i.e.

$$\frac{\Delta\beta}{\beta} \ll \frac{w}{l_c}$$

The precision required to control elements for compensating tune spread $\frac{\Delta\beta}{\beta}$ is

$$\frac{w}{l_c}\frac{\beta}{\Delta\beta}.$$

Random linear optics errors lead to a requirement for beta function control:

$$\frac{\Delta\beta}{\beta} \ll \frac{w}{l_c}\sqrt{\frac{l_c}{\beta}}.$$

## VI. PIC IN A SNAKE CHANNEL

### A. Alternating dispersion and linear optics design

The main constraint of parametric resonance ionization cooling channel design is to combine low dispersion at the wedge absorber plates (for emittance exchange to compensate energy straggling) with large dispersion in the space between plates (where sextupoles can be placed to compensate for chromatic aberration). This constraint can be achieved in a channel created by an alternating dipole field as indicated in figure 5.

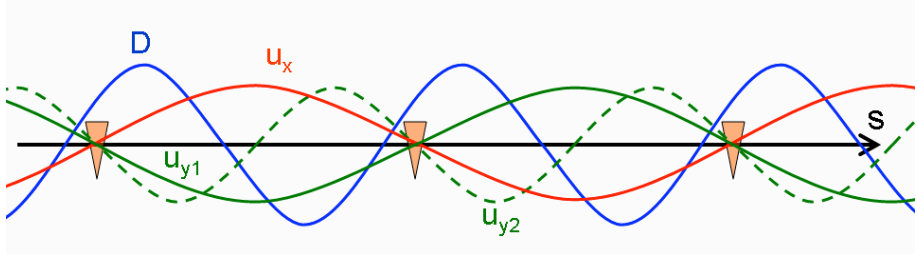

**Figure 5:** Lattice functions of a beam cooling channel suitable for PIC showing D, the dispersion (blue), the horizontal betatron amplitude $u_x$ (red), and two possible solutions for the vertical betatron amplitude, $u_{y1}$ and $u_{y1}$ (green). The triangles represent the wedge absorbers. The dipoles and quadrupoles are not shown.

In such a channel, the dispersion alternates with the orbit displacement. The absorber plates then are positioned near zero dispersion points. Obviously, the betatron (i.e. focusing) wave length in horizontal plane must not be the same as the bend and dispersion periods, but it should be two times longer. Thus one should design the alternating bend, non-coupled linear optics channel such that the dispersion and wedges maintain the minimum momentum spread (~2.5%) and that the dispersion is small enough at the absorbers to prevent a *straggling impact* on the horizontal emittance, but large in between to compensate for *chromatic detuning*.

The notation for the following discussion follows the normal conventions:

$$x'' + (K^2 - n)x = Kq$$

$y'' + ny = 0$

$$x = Dq + \tilde{x}$$
$$x' = D'q + \tilde{x}'$$

$$D'' + (K^2 - n)D = K$$
$$\tilde{x}'' + (K^2 - n)\tilde{x} = 0$$

$$\tilde{x} = a_x u_x + b_x v_x$$
$$\tilde{x}' = a_x u'_x + b_x v'_x$$

$$y = a_y u_y + b_y v_y$$
$$y' = a_y u'_y + b_y v'_y$$

$$u_{x,y}(-s) = u_{x,y}(s)$$
$$v_{x,y}(-s) = -v_{x,y}(s)$$

Three possible cases suitable for PIC are:

$$\lambda_x = \lambda_y \qquad (K^2 - n \approx n)$$

$$\lambda_y = 2\lambda_x \qquad (K^2 - n = 4n)$$

$$\lambda_x = 2\lambda_y \qquad (K^2 - n \approx \frac{n}{4})$$

To avoid resonance between the dispersion and horizontal betatron motion, the dispersion period $\lambda_0$ must be half or one quarter of $\lambda_x$. The third option seems preferred due to large horizontal aperture.

It is convenient to use a symmetric lattice to simplify the linear optics design and the addition of aberration correction elements. The symmetry relative to the absorbers can be chosen naturally with $K$ and $D$ antisymmetric and $K^2$ and $n$ symmetric. Note that $D$ is not exactly antisymmetric since some small dispersion is needed at the absorbers in order to create the emittance exchange required for longitudinal cooling.

Since the lattice we need is designed with correlated dispersion and betatron motion, we are sensitive to unwanted strong linear structure parametric resonances that must be avoided.
The desired driving parametric resonance can be arranged in two planes by modulation of $B$ and $n$, but preferably by specific weak lenses.

### B. Parametric resonance in a horizontal plane

Here we derive the equations for a perfectly tune snake channel with no tune spread. Consider motion in two planes:
$$x'' + k^2(1 + \Delta_x)^2 x - 4\zeta k^2 x \sin 2ks = Kq$$

$$y'' + 4k^2(1 + \Delta_y)^2 y + 4\zeta k^2 y \sin 2ks = 0$$

where $(\Delta_y, \Delta_x) \ll 1$ are constant detunes in two planes from exact correlated optics, while $\zeta = const \ll 1$ is frequency modulation parameter due to introduced alternated quadrupole field.

Let us represent the oscillator motion in terms of "slow" variables $a(z), b(z)$ that would be constant at detunes and modulation

$$x = Dq + x_b = Dq + a_x \cos ks + b_x \sin k s$$
$$x' = D'q + x'_b = D'q + k(-a_x \sin k s + b_x \cos ks)$$

$$y = a_y \cos 2ks + b_y \sin 2ks$$
$$y' = 2k(-a_y \sin 2ks + b_y \cos 2ks)$$

Here, $a$ and $kb$ represent particle coordinate and angle in the x-plane, respectively, at points of the beam orbit where $\sin kz = 0$, while at points where $\cos kz = 0$ the coordinate and angle are correspondently represented by b and $ka$. Let us rewrite above relationships using the variables $a$ and $b$ as functions of $x$ and $x'$:

$$a_x = (x - Dq) \cos ks - \frac{1}{k}(x' - D'q) \sin ks$$
$$b_x = (x - Dq) \sin ks + \frac{1}{k}(x' - D'q) \cos ks$$

$$a_y = y \cos 2ks - \frac{1}{2k} y' \sin 2ks$$
$$b_y = y \sin 2k s + \frac{1}{2k} y' \cos 2ks$$

$$x'' + k^2(1 + \Delta_x)^2 x - 4\zeta k^2 x \sin 2ks = Kq$$

$$y'' + 4k^2(1 + \Delta_y)^2 y + 4\zeta k^2 y \sin 2ks = 0$$

Then, we can easily find the derivatives $a'$ and $b'$ by taking into account the equation of motion:

$$a'_x = 2k\Delta_x x \sin ks - 4\zeta k x \sin ks \cdot \sin 2ks$$
$$b'_x = -2k\Delta_x x \cos ks + 4\zeta k x \cos ks \cdot \sin 2ks$$

$$a'_y = 4k\Delta_y y \sin 2ks + 2k\zeta y \sin^2 2ks$$
$$b'_y = -4k\Delta_y y \cos 2ks - k\zeta y \sin 4ks$$

Thus, we derived the equations of motion in terms of slow variables $a$ and $b$. The coefficients of these equations are periodic with period $\pi/k$. Assuming that both parameters $\zeta$ and $v$ are small, we can approximate the change of $a$ and $b$ during a single period by simply integrating the equations over one period at a =const, b = const on the right side of the equations. Then we find:

$$a'_x \Rightarrow \Delta_x k b_x - \zeta k a_x$$
$$b'_x \Rightarrow -\Delta_x k a_x + \zeta k b_x$$

$$a'_y \Rightarrow 2\Delta_y k b_y$$
$$b'_y \Rightarrow -2\Delta_y k a_y$$

At ideal tuning $\Delta_x = 0$
$$a'_x = -\zeta k a_x$$
$$b'_x = \zeta k b_x$$

$$a_x = a_x(0)\exp(-\zeta ks)$$
$$b = b_x(0)\exp(\zeta ks)$$

Let us assume $\zeta > 0$. According to these equations the beam experiences an exponential shrinkage at points where $\sin kz = 0$, where we can put absorbers. The angle spread would grow exponentially if it were not constrained by the ionization cooling due to the absorber. Correspondently, the beam size grows while angle spread damps at points $\cos kz = 0$. Thus, at the exact parametric resonance we observe the hyperbolic dynamics identical to that described in section II above.

At a finite detune,
$$\begin{vmatrix} \kappa + \zeta, & \Delta_x \\ -\Delta_x, & \kappa - \zeta \end{vmatrix} = 0$$

$$\kappa = \pm\sqrt{\zeta^2 - \Delta_x^2}$$

By complementing the lattice with skew quadrupole harmonic $\propto \sin 4ks,$ it is possible to have the parametric resonance condition also in the vertical plane, to reach the beam focused simultaneously at the same point in both planes.

## C. Two-dimensional parametric resonance by introducing a coupling resonance

Another possibility that could be more convenient would be to introduce a coupling resonance as a way to equalize the cooling rates in each transverse plane and to simplify the parametric resonance design by using skew-quads.

Other important attribute of the coupling resonance is that it also can be effectively used to reduce the number of compensation conditions for detuning due to numerous chromatic, geometrical, and non-linear aberrations in the optics of the PIC channel. This reduction becomes critical while approaching equilibrium at the end of the PIC process.

*Coupling resonance in correlated optics channel*

It is critically important to design the two betatron periods to be different by a factor of 2 so that the coupling resonance causes beam rotation in space at the absorber, while keeping the horizontal and vertical beam sizes small and maintaining the PIC condition. The skew gradient field should alternate with a frequency equal to the difference of the two betatron frequencies.

$$x'' + k^2 x = Kq + 4gk^2 y \sin ks$$

$$y'' + 4k^2 y = 4gk^2 x \sin ks$$

$$x_b'' + k^2 x_b = -4gk^2 y \sin ks$$

$$y'' + 4k^2 y = -4gk^2 (Dq + x_b) \sin ks$$

$$a'_x = 4gky \sin ks \cdot \sin ks$$
$$b'_x = -4gky \sin ks \cdot \cos ks$$

$$a'_y = 2gk \sin 2ks \cdot (Dq + x_b) \sin ks$$
$$b'_y = -2 gk\cos 2ks \cdot (Dq + x_b) \sin ks$$

By applying the averaging method to the equations of $x$ and $y$ motion in terms of slow variables $a$ and $b$ as above used in the case of the parametric resonance, we obtain the coupled x-y dynamics with equations as follows:

$$a'_x = -gka_y; \qquad a'_y = \frac{1}{2}gka_x$$

$$b'_x = -gkb_y; \qquad b'_y = \frac{1}{2}gk\, b_x$$

$$a''_x + \frac{1}{2}(gk)^2 a_x; \qquad a''_y + \frac{1}{2}(gk)^2 a_y;$$

$$b''_x + \frac{1}{2}(gk)^2 b_x; \qquad b''_y + \frac{1}{2}(gk)^2 b_y;$$

$$\Omega = gk/\sqrt{2}$$

$$a_x + i\sqrt{2}a_y = A\exp(i\Omega s)$$
$$b_x + i\sqrt{2}b_y = B\exp(i\Omega s)$$

$$a_x = |A|\cos \Phi_a; \quad a_y = \frac{1}{\sqrt{2}}|A|\sin \Phi_a;$$
$$b_x = |B|\cos \Phi_b; \quad b_y = \frac{1}{\sqrt{2}}|B|\sin \Phi_b;$$
$$\Phi'_a = \Phi'_b = \Omega$$

The equations show that the coupling resonance causes beam rotation at the absorbers (where the beam is focused by the parametric resonance), while it does not change the sum of the Courant-Snyder actions.

$$a_x^2 + 2a_y^2 = |A|^2 = const$$
$$b_x^2 + 2b_y^2 = |B|^2 = const$$

*PIC channel with parametric and coupling resonance*

$$a'_x = \Delta_x k b_x - \zeta k a_x - gka_y$$
$$a'_y = 2\Delta_y k b_y + \frac{1}{2}gka_x$$

$$b'_x \Rightarrow -\Delta_x k a_x + \zeta k b_x - gkb_y;$$
$$b'_y \Rightarrow -2\Delta_y k a_y + \frac{1}{2}gk\, b_x$$

$$A = (a_x + i\sqrt{2}a_y)\exp(-i\Omega s)$$
$$B = (b_x + i\sqrt{2}b_y)\exp(-i\Omega s)$$

$$A' = (a_x + i\sqrt{2}a_y)'\exp(-i\Omega s) - i\Omega A$$
$$B' = (b_x + i\sqrt{2}b_y)'\exp(-i\Omega s) - i\Omega B$$

$$A' == k(\Delta_x b_x + 2i\sqrt{2}\Delta_y b_y - \zeta a_x)\exp(-i\Omega s)$$
$$B' = k(-\Delta_x a_x - 2i\sqrt{2}\Delta_y a_y + \zeta b_x)\exp(-i\Omega s)$$

If the coupling parameter *g* is large enough compared to the rates of the parametric resonance and ionization cooling, the coupling resonance will effectively equalize the PIC dynamics and cooling rates of the two planes.

$$A' \Rightarrow -\frac{\zeta}{2}kA + (\frac{1}{2}\Delta_x + \Delta_y)kB$$

$$B' \Rightarrow \frac{\zeta}{2}kB - (\frac{1}{2}\Delta_x + \Delta_y)kA$$

### D. 2d PIC in snake channel with coupling resonance

$$A' = k[\Delta_x b_x + 2i\sqrt{2}\Delta_y b_y - \zeta a_x + (a_x + i\sqrt{2}a_y)'_i]\exp(-i\Omega s)$$

$$B' = k[-\Delta_x a_x - 2i\sqrt{2}\Delta_y a_y + \zeta b_x + (b_x + i\sqrt{2}b_y)'_i]\exp(-i\Omega s)$$

$$(a'_y)_i = -(y'')_i \frac{\sin 2ks}{2k}$$

$$(b'_y)_i = (y'')_i \frac{\cos 2ks}{2k}$$

$$(a'_x)_i = -(x'')_i \frac{\sin ks}{k} - \frac{(\gamma')_i}{\gamma v^2}(D \cos ks - D' \frac{\sin ks}{k})$$

$$(b'_x)_i = (x'')_i \frac{\cos ks}{k} - \frac{(\gamma')_i}{\gamma v^2}(D \sin ks + D' \frac{\cos ks}{k})$$

$$(\gamma')_i = \overline{(\gamma')_i} + (\gamma')_{sc}$$

$$(x'')_i = \frac{\overline{(\gamma')_i}}{\gamma v^2}x' + (x'')_{sc}$$

$$(y'')_i = \frac{\overline{(\gamma')_i}}{\gamma v^2} y' + (y'')_{sc}$$

$\delta |A|^2 = A^* \delta A + A \delta A^* + |\delta A|^2$

$\delta |B|^2 = B^* \delta B + B \delta B^* + |\delta B|^2$

$<|A|^2>' = A^* <A'> + A <A'^*> + <|\delta A|^2>'$

$<|B|^2>' = B^* <B'> + B <B'^*> + <|\delta B|^2>'$

$D'_a = 0$

$A' = [k(\Delta_x b_x + 2i\sqrt{2}\Delta_y b_y - \zeta a_x) + (a_x + i\sqrt{2}a_y)'_i]\exp(-i\Omega s)$

$B' = [-k(\Delta_x a_x + 2i\sqrt{2}\Delta_y a_y - \zeta b_x) + (b_x + i\sqrt{2}b_y)'_i]\exp(-i\Omega s)$

$<A'> = \left(\frac{\Delta_x}{2} + \Delta_y\right) kB + \frac{1}{2}\left(-k\zeta + \frac{\Lambda_6}{v^2}\frac{D_a}{h}\right) A \equiv \left(\frac{\Delta_x}{2} + \Delta_y\right) kB - \Lambda_A A$

$<B'> = -\left(\frac{\Delta_x}{2} + \Delta_y\right) kA + \frac{1}{2}\left(k\zeta - \frac{\Lambda_6}{v^2}\right) B \equiv -\left(\frac{\Delta_x}{2} + \Delta_y\right) kA - \Lambda_B B$

$<\Delta\gamma'> = \frac{\Lambda_6}{v^2}\left(\frac{1}{\gamma^2} - \frac{D_a}{2h}\right)\Delta\gamma \equiv -\Lambda_\gamma \Delta\gamma$

$\Lambda_6 \equiv -2\frac{<(\gamma')_i>}{\gamma}$

$2(\Lambda_A + \Lambda_B) + \Lambda_\gamma = \Lambda_6$

$2\Lambda_A = 2\Lambda_B = \Lambda_\gamma = \frac{1}{3}\Lambda_6$

$$\Lambda_\gamma = \frac{1}{3}\Lambda_6 \rightarrow D_a = 2h(1 - \frac{2}{3}v^2)$$

$\zeta = \frac{\Lambda_6}{2kv^2}(3 - \frac{4}{3}v^2)$

$<|\delta A|^2>' = <(\delta a_x)^2 + 2(\delta a_y)^2>'$

$<|\delta B|^2>' = <(\delta b_x)^2 + 2(\delta b_y)^2>'$

$\delta a_y = -\delta y' \frac{\sin 2ks}{2k}; \quad \delta a_x = -\delta x' \frac{\sin ks}{k} - \frac{\delta p}{p} D_a \cos ks$

$$\delta b_y = \delta y' \frac{\cos 2ks}{2k}; \quad \delta b_x = \delta x' \frac{\cos ks}{k} - \frac{\delta p}{p} D_a \sin ks$$

$$<|\delta A|^2>' = <2(\delta y' \frac{\sin 2ks}{2k})^2>' + <(\delta x' \frac{\sin ks}{k})^2>' + <(\frac{\delta p}{p} D_a \cos ks)^2>'$$

$$<|\delta B|^2>' = = <2(\delta y' \frac{\cos 2ks}{2k})^2>' + (\delta x' \frac{\cos ks}{k})^2>' <(\frac{\delta p}{p} D_a \sin ks)^2>'$$

$$<|\delta A|^2>' = \frac{m_e}{m_\mu \gamma^2 v^2} <|\gamma_i'|[\frac{Z+1}{k^2}(\frac{1}{2}\sin^2 2ks + \sin^2 ks) + \frac{\gamma^2+1}{2\log}(D_a \cos ks)^2]>$$

$$<|\delta B|^2>' = \frac{m_e}{m_\mu \gamma^2 v^2} <|\gamma_i'|[\frac{Z+1}{k^2}(\frac{1}{2}\cos^2 2ks + \cos^2 ks) + \frac{\gamma^2+1}{2\log}(D_a \sin ks)^2]>$$

$$<|\delta A|^2>' \Rightarrow \Lambda_6 \frac{m_e}{4m_\mu \gamma v^2}(\frac{Z+1}{2}w^2 + \frac{\gamma^2+1}{\log}D_a^2)$$

$$<|\delta B|^2>' \Rightarrow \Lambda_6 \frac{3m_e}{4m_\mu \gamma v^2}\frac{Z+1}{k^2}$$

$$<|A|^2>' = -\frac{\Lambda_6}{3}<|A|^2> + <|\delta A|^2>'$$

$$<|B|^2>' = -\frac{\Lambda_6}{3}<|B|^2> + <|\delta B|^2>'$$

$$<|A|^2> \Rightarrow <|A|^2>_{eq} = \frac{3m_e}{4m_\mu \gamma v^2}(\frac{Z+1}{2}w^2 + \frac{\gamma^2+1}{\log}D_a^2)$$

$$<|B|^2> \Rightarrow <|B|^2>_{eq} = \frac{9m_e}{8m_\mu \gamma v^2}\frac{Z+1}{k^2}$$

$$<a_x^2> = 2<a_y^2> = \frac{1}{2}<|A|^2>$$

$$<\theta_x^2> = k^2<b_x^2> = 2k^2<b_y^2> = \frac{1}{2}<\theta_y^2> = \frac{k^2}{2}<|B|^2>$$

$$\epsilon_x = \sqrt{<a_x^2><\theta_x^2>} = \frac{k}{2}\sqrt{<|A|^2><|B|^2>}$$

$$\epsilon_y = \epsilon_x$$

$$\epsilon_n = \gamma v \epsilon \Rightarrow \frac{3\sqrt{3}(Z+1)}{16v} \frac{m_e}{m_\mu} w \sqrt{(1 + \frac{2}{\log} \frac{\gamma^2+1}{Z+1} \frac{D_a^2}{w^2})}$$

$$<|A|^2>' = \left(\frac{\Delta_x}{2} + \Delta_y\right) k <(A^*B + AB^*)> - \frac{\Lambda_6}{3} <|A|^2> + <|\delta A|^2>'$$

$$<|B|^2>' = -Ck <(A^*B + AB^*)> - \frac{\Lambda_6}{3} <|B|^2> + <|\delta B|^2>'$$

$$<A'> = +\frac{1}{2}\left(-k\zeta + \frac{\Lambda_6}{v^2}\frac{D_a}{h}\right) A \equiv \left(\frac{\Delta_x}{2} + \Delta_y\right) kB - \Lambda_A A$$

$$<B'> = -\left(\frac{\Delta_x}{2} + \Delta_y\right) kA + \frac{1}{2}\left(k\zeta - \frac{\Lambda_6}{v^2}\right) B \equiv -\left(\frac{\Delta_x}{2} + \Delta_y\right) kA - \Lambda_B B$$

$$<(A^*B + AB^*)>' = -2k\left(\frac{\Delta_x}{2} + \Delta_y\right)(<|B|^2 - |A|^2>) - \frac{\Lambda_6}{3} <(A^*B + AB^*)>$$

$$<(A^*B + AB^*)> = -\frac{6k}{\Lambda_6}\left(\frac{\Delta_x}{2} + \Delta_y\right)(<|B|^2 - |A|^2>)$$

at

$$k\left(\frac{\Delta_x}{2} + \Delta_y\right) \ll \frac{\Lambda_6}{6},$$

$$<|A|^2> \approx <|A|^2>_0 + <|B|^2> \frac{1}{2}\left[\frac{6k}{\Lambda_6}\left(\frac{\Delta_x}{2} + \Delta_y\right)\right]^2$$

so, it should be:
$$\left(\frac{\Delta_x}{2} + \Delta_y\right) \ll \frac{\Lambda_6}{6} w$$

An example of a particular implementation of the snake channel is presented in Appendix B.

## VII. COMPENSATION FOR ABERRATIONS IN SNAKE CHANNEL

The realization of Parametric-resonance Ionization Cooling requires compensation for tunes spreads caused by numerous aberrations: chromatic, spherical, and non-linear field effects. Our analysis of the compensation is based on the expansion of Hamiltonian function for a plane snake type beam transport derived in the Appendix A.

### A. Compensation for the chromatic effects

*The 3d power Hamiltonian terms*

The 3d order terms to be compensated are seen in the expansion of the Hamiltonian presented in

Appendix:

$$H_3 = -\frac{1}{2}(p_x^2 + p_y^2)(q - Kx) - \frac{1}{2}K[(K^2 + n)x^3 - nxy^2]$$

$$- \alpha_3(x^3 + 3xy^2) - K'xyp_y - \frac{1}{3}n_{sext}(x^3 - 3xy^2)$$

$$\alpha_3 \equiv \frac{1}{12}(K'' - 3K^3 - Kn)$$

where the parameter $n_{sext}$ is the field index of the introduced sextupole components.

*Chromatic Hamiltonian*

The chromatic terms in the initial general form are defined as linear on particle momentum deviation from the reference particle:
$H_3 \Longrightarrow -q\mathfrak{H}_3$;
here

$$\mathfrak{H}_3 = a_x^2 U_x + b_x^2 V_x + a_x b_x W_x + a_y^2 U_y + b_y^2 V_y + a_y b_y W_y$$

$$U_x \equiv \frac{1}{2}(1 - KD){u'}_x^2 + \left[\frac{3}{2}K(K^2 + n) + 3\alpha_3 + n_{sext}\right]Du_x^2 - KD'u_x u'_x$$

$$V_x \equiv \frac{1}{2}(1 - KD){v'}_x^2 + \left[\frac{3}{2}K(K^2 + n) + 3\alpha_3 + n_{sext}\right]Dv_x^2 - KD'v_x v'_x$$

$$U_y \equiv \frac{1}{2}(1 - KD){u'}_y^2 + \left(3\alpha_3 - \frac{1}{2}Kn - n_{sext}\right)Du_y^2 + K'Du_y u'_y$$

$$V_y \equiv \frac{1}{2}(1 - KD){v'}_y^2 + \left(3\alpha_3 - \frac{1}{2}Kn - n_{sext}\right)Dv_y^2 + K'Dv_y v'_y$$

$$W_x \equiv (1 - KD)u'_x v'_x - KD'(u_x v_x)' + [3K(K^2 + n) + 6\alpha_3 + 2n_{sext}]Du_x v_x$$

$$W_y \equiv (1 - KD)u'_y v'_y + (6\alpha_3 - Kn - 2n_{sext})Du_y v_y + K'D(u_y v_y)'$$

Rate of change of amplitudes $a$ and $b$ is given by the canonical equations:

$$b'_x = \frac{\partial H_3}{\partial a_x} = (2U_x a_x + W_x b_x)q\,;$$

$$a'_x = -\frac{\partial H_3}{\partial b_x} = -(2V_x b_x + W_x a_x)q\,;$$

and similar for the $y$ plane.

In our *correlated optics* PIC channel, all the coefficient functions of the Hamiltonian are periodic on the longitudinal coordinate $s$ with period of lattice or twice of it. Therefore, compensation for aberrations due to chromatic and non-linear effects are equivalent to demand that change of amplitudes $a$ and $b$ over a period should be equal to zero or small enough in order not to disturb significantly the ionization cooling process compared to the case of an ideal linear optics tuned to the parametric resonance. Change for a period can be found by integration of equations for the amplitudes a and b over a single period, using the iteration method. In the first order approximation, the change is given by integration of the right sides over a period at constant amplitudes $a$ and $b$; so the compensation requirement can be express in terms of averaging over a period:

$$H_3 \Rightarrow <H_3> = <q\mathfrak{H}_3> \Rightarrow 0.$$

*Compensation for chromatic tune spread*

Let us first consider the dynamics in absence of absorber and RF field, i.e. $q = const$.

$q = const$

$<H_3> \Rightarrow q <\mathfrak{H}_3>;$

$<\mathfrak{H}_3> = a_x^2 <U_x> + b_x^2 <V_x> + a_y^2 <U_y> + b_y^2 <V_y> + a_x b_x <W_x> + a_y b_y <W_y>$

The compensation for chromaticity problem is simplified for a lattice which is symmetric about the mid-point between absorbers; this conditioning can be naturally arranged. Taking this symmetry into account, we find:

$$U_x(s) = U_x(-s), \quad V_x(s) = V_x(-s),$$

and similar for $U_y$ and $V_y$, while:

$$W_x(s) = -W_x(-s), \quad W_y(s) = -W_y(-s).$$

So we have
$<W_x> = 0, \quad <W_y> = 0.$

Then, the averaged Hamiltonian function $<H_3>$ is reduced to the following expression:

$<\mathfrak{H}_3> = a_x^2 <U_x> + b_x^2 <V_x> + a_y^2 <U_y> + b_y^2 <V_y>;$

and the correspondent averaged equations for the amplitudes are as follows:

$b'_x = -2q a_x <U_x>$

$a'_x = 2q b_x <V_x>$

*Complete compensation for the linear chromaticity*

There are four equations for complete linear compensation of chromatic tunes:

$<U_x> = <V_x> = <U_y> = <V_y> = 0$

*Sufficient compensation*

In practice, it might be enough to satisfy only two requirements

$$< U_x > = < U_y > = 0,$$

thus removing excitation of the $v$-modes of particle trajectories by the chromatic detunes. In this way, the introduced parametric resonance will provide beam focusing at the absorbers, consequently reducing the scattering effect. Estimates for typical momentum spreads show that, perturbation of the $u$-modes according to equation for the amplitude $a$ (i.e. increase of angle spread at absorbers) is relatively small.

*Design of compensating sextupoles*

To satisfy compensation conditions, the compensating sextupole field should be designed for case $\lambda_x = 2\lambda_y = 4\lambda_0$ reflecting the field behavior along the beam path by two harmonics as follows:

$$n_s(s) \propto n_{s1} \sin k_0 s + n_{s2} \sin 2k_0 s \; ; \quad k_0 = 2\pi/\lambda_0$$

*Parasitic parametric resonance and its compensation*

With absorbers and RF field in the channel, particle energy in a single period varies with coordinate $s$. Considering integration of equations for the amplitudes inside a single cell, one can represent variable $q$ as
$$q = \bar{q} + \tilde{q} \; ; \quad \tilde{q}(s) = -\tilde{q}(-s); \quad <\tilde{q}> = 0$$

where $\bar{q}$ is constant inside a period, but is experiences a change ("jump") over every period with particle synchrotron phase $\varphi$ (in linear approximation):

$$\Delta \bar{q} = V_a \varphi.$$

The alternate part $\tilde{q}$ is phase dependent together with $\Delta\bar{q}$:

$$\tilde{q}(s, \varphi) = \tilde{q}_0(s) + \tilde{q}_1(s)\varphi.$$

The average Hamilton function now has a "parasitic" parametric resonance addition, due to that $\tilde{q}$ oscillates with a period twice (and four times) shorter compared to the periods of the $y$ and $x$ betatron oscillation, respectively:

$$H_{ppr} = -<\tilde{q}\mathfrak{H}_3> = -<\tilde{q}W_x> a_x b_x - <\tilde{q}W_y> a_y b_y.$$

Note that,
$$<\tilde{q}U_x> = <\tilde{q}V_x> = <\tilde{q}U_y> = <\tilde{q}V_y> = 0$$

due to the odd symmetry of $\tilde{q}(s)$.

The change in rates of *b* and *a* are given by the canonical equations:

$$b'_x = \frac{\partial H_{pr}}{\partial a_x} = -<\tilde{q}W_x> b_x ;$$

$$a'_x = -\frac{\partial H_{pr}}{\partial b_x} = <\tilde{q}W_x> a_x ;$$

here

$$<\tilde{q}W_x> = <\tilde{q}_0 W_x> + <\tilde{q}_1 W_x>\varphi .$$

Strength of this parasitic resonance is of the order of value of cooling decrement, hence about same strength as of the induced, the driving parametric resonance. The phase independent part of the parasitic resonance can be compensated by a simple correction of the resonance driving force. Effect of the phase dependent part is relatively small due to the two factors. One is that phases $\varphi$ particles are small. Other is that $\varphi$ are experiencing the synchrotron oscillations, frequency of which is large compared to the cooling and parametric resonance decrements. After all, phase dependent part can also be compensated by a specific RF arrangement, if needed; however, such sophistication seems not required.

### B. Compensation for spherical and other 4th order aberrations

*The 4$^{th}$ order Hamiltonian terms*

The spherical and some other geometrical terms to be compensated are seen in the fourth power terms of the expansion of the Hamiltonian:

$$H_4 = \frac{1}{8}(p_x^2 + p_y^2)^2 + \frac{1}{2}(K'xy)^2 - \alpha_3 K(x^4 + 3x^2 y^2) - \alpha_4(x^4 - y^4)$$
$$- \beta_4 x^2 y^2 - \frac{1}{3}Kn_{sext}(x^4 - 3x^2 y^2) - \frac{1}{4}n_{oct}(x^4 + y^4 - 6x^2 y^2)$$

Here we use the following notation:

$$\alpha_3 \equiv \frac{1}{12}(K'' - 3K^3 - Kn)$$
$$\alpha_4 = \frac{1}{24}(K^2 n - n'' + \delta)$$
$$\beta_4 = -\frac{3}{2}K\alpha_3 + \frac{1}{4}\delta$$
$$\delta = \frac{9}{2}K^4 + K^2 n - 2K'^2 - 3KK'' ,$$

and $n_{oct}$ is the field index of the introduced octupole components.

*Compensation for the 4$^{th}$ order effects*

The main compensation conditions that are connected to the 4th power of oscillation mode are antisymmetric about the absorber points of the snake orbit. (Note that the symmetric oscillation components damp!) These conditions are:

$$8 < (K\alpha_3 + \alpha_4 + \frac{1}{3}Kn_{sext} + \frac{1}{4}n_{oct})u_x^4 > = < u_x'^4 >$$

$$8 < (\frac{1}{4}n_{oct} - \alpha_4)u_y^4 > = < u_y'^4 >$$

$$4 < (3K\alpha_3 + \beta_4 - Kn_{sext} - \frac{3}{2}n_{oct})(u_x u_y)^2 > = < (u_x' u_y')^2 >$$

The compensating octupole field design for the case that $\lambda_x = 2\lambda_y = 4\lambda_0$ should include a constant component and the two lowest harmonics of the lattice period:

$$n_{oct} = n_{0oct} + n_{1ont}\cos k_0 s + n_{2oct}\cos 2k_0 s.$$

- Typical magnitudes of the spherical and non-linear detunes
- Coupling resonance and non-linear instabilities in a channel with correlated optics
- Required precision control of the compensating magnets

*Compensation for Higher Order Aberrations*

Other 4$^{th}$ power terms result from the second order effect of the 3d power Hamiltonian as shown in the Appendix. Finally, there are several higher power terms in the expansion of the Hamiltonian function which may also require compensation. Compensation for all these terms can be greatly simplified by the use of a coupling resonance, as discussed above. The relevant compensation tools (related multipole field components, beam simulations, and experimental diagnostics) are subject of further study. One example involves simulation software COSY INFINITY [9] that is a highly efficient tool for aberration corrections (Appendix C).

### VIII. TUNE SPREAD DUE TO MUON SPACE CHARGE IN THE PIC CHANNEL

The tune spread due to the space charge of muon bunch under PIC can be estimated from first principles based on:

$$x'' + k^2 x = f_x \qquad (11)$$

$$k^2 \equiv \frac{1}{\beta^2}$$

with space charge transverse force, $f_x$. The maximum tune shift is produced for particles near the beam center, while for large amplitude particles (beam tails) the shift tends to zero. Therefore, the absolute tune spread value can be found by calculating the tune shift for the center particles. Here, assuming bunch longitudinal size large compared to the transverse beam size (round beam), the transverse space charge force can be approximated by linear behavior as follows:

$$f_x \Rightarrow f_0 \equiv \frac{2\pi n_0(s) r_\mu}{\gamma^3 v^2} x,$$

where the density factor $n_0(s)$ is controlled by the beam envelope developed in the PIC process. For a Gaussian charge distribution in the bunch, the density factor is found equal to

$$n_0(s) = \frac{N}{(2\pi)^{3/2} \sigma_\perp^2(s) \sigma_s}$$

with $\sigma_\perp(s)$ and $\sigma_s = const$ for beam rms size in transverse and longitudinal direction, respectively. Since the tune spread due to space charge is supposed to be small (as for any source of tune spread), the effect can be calculated by the perturbation method. For this, we transform the description of transverse dynamics to variables $a$ and $\psi$, according to the representation

$$x = a \sin \psi \qquad (12)$$
$$x' = ka \cos \psi \qquad (13)$$

For unperturbed motion in a focusing lattice, we have $a = const$ and $\psi = ks + const$. Our aim is to calculate the average change of phase advance per betatron period due to the space charge force. For this, we express the phase $\psi$ as a function of $x$ and $x'$ using equations (12) and (13):

$$tg\psi = \frac{kx}{x'}.$$

By taking the derivative of this equation along "time" $s$, we find a relationship

$$\frac{\psi'}{\cos^2 \psi} = k - \frac{kxx''}{x'^2};$$

By substituting $\cos\psi$ according to equation (11) and $x''$ according to equation (13) we find:

$$\psi' = k - \frac{fx}{ka^2}.$$

Now, we will integrate this equation along the unperturbed oscillation trajectory $a = const$, $\psi = ks$, for a particle near the beam center:

$$(\delta\psi')_0 = -\frac{f_0 x}{ka^2} = -\frac{2\pi n_0 r_\mu}{\gamma^3 v^2 k} \frac{x^2}{a^2}$$

or

$$(\delta\psi')_0 = -\frac{Nr_\mu \beta}{\gamma^3 v^2 \sqrt{2\pi}\sigma_z} \frac{\sin^2 \psi}{\sigma_\perp^2(s)}.$$

In the PIC process, particle trajectories tend to focus at absorber points, then there is a full spread in amplitudes $\Delta a = \sigma_0 \Rightarrow \beta\theta^*$ determined by the angle spread at the absorbers

$$\theta^{*2} = \frac{3(Z+1)m_e}{2\gamma v^2 m_\mu} \qquad (14)$$

Thus, the beam envelope at the end of the PIC process is described by

$$\sigma_\perp^2(s) \Rightarrow \sigma_0^2 \sin^2 ks + \sigma^{*2}, \quad \sigma_0^2 \gg \sigma^{*2}.$$

Then,

$$-(\delta\psi')_0 \Rightarrow \frac{Nr_\mu \beta}{\gamma^3 v^2 \sigma_z \sqrt{2\pi}} \frac{\sin^2 ks}{\sigma_0^2 \sin^2 ks + \sigma^{*2}} < \frac{Nr_\mu \beta}{\sigma_z \sqrt{2\pi}\gamma^3 v^2 \sigma_0^2}$$

Note, that the moment of crossing the minimum transverse size (at the absorber) does not bring a dominating contribution to the space charge phase advance. In fact, the instantaneous tune shift is almost constant along the beam envelope. Then we find an accurate estimation for phase advance per period of betatron oscillation and tune shift for center particles, and tune spread:

$$(\delta\psi)_0 = \oint (\delta\psi') ds \approx 2\pi\beta(\delta\psi')_0$$

$$\Delta\nu = -\delta\nu_0 \equiv -\frac{(\delta\psi)_0}{2\pi} \approx \frac{Nr_\mu}{\gamma^3 v^2 \sigma_z \sqrt{2\pi}} \frac{\beta^2}{\sigma_0^2} = \frac{Nr_\mu}{\gamma^3 v^2 \theta^{*2} \sigma_z \sqrt{2\pi}}$$

Finally, by taking into account relationship (14), we find:

$$\Delta\nu = \frac{2Nr_e}{3(Z+1)\gamma^2 \sigma_z \sqrt{2\pi}}.$$

This formula was used for the numerical estimate of tune spread due to the space charge at PIC equilibrium shown in Table 2.

It should be noted that the tune spread due to space charge is determined in PIC not by the beam transverse phase space (true emittance reduced by PIC), but by the beam characteristic size between absorbers and beam optics along the PIC channel. Remarkably, the tune spread is determined, in fact, simply by the angle spread at the absorbers, which is not function of beam optics at all. Finally, the resulting expression for tune spread depends only on the absorber atomic number Z. The beam transverse phase space (true emittance) after PIC will be transformed by matching optics to a conventionally defined beam emittance.

## IX. AN OPTIMIZED PIC DESIGN

In order to approach a practical PIC design, the cooling channel parameters can be modified as the beam is cooled.

First of all and in general, PIC may start with reasonably thick absorbers in order to minimize the beam path while developing the parametric resonance beam envelope.

The maximum initial cooling rate may be limited by the available accelerating RF power. To alleviate this limitation, the initial cooling sections can be effectively isochronous, without RF, where the lattice magnet strengths scale with the decreasing beam momentum. The beam energy could then be restored by RF cavities installed between cooling sections.

Also note that at the beginning and during the middle PIC stages, high precision compensation for aberrations is not required since the beam is not so well focused. Thus the design of the PIC channel for the initial stages can be relatively easy compared to the final stage.

## X. CONCLUSIONS

In this paper we have introduced the basic theory of ionization cooling using parametric resonances and discussed the requirements that must be satisfied to achieve significant beam cooling using this technique. We have derived the effects of chromatic, spherical, and non-linear field aberrations and described how these aberrations can be compensated. That PIC must be accompanied by emittance exchange and that these two techniques are compatible has been demonstrated. We have suggested a particular transport scheme using alternating bends to achieve the two requirements of small dispersion at absorbers and large dispersion where aberrations can be compensated. In addition, the use of a conservative coupling resonance is proposed with the purpose of providing equal parametric resonance and cooling rates in the two planes of the beam transport line, and of simplifying the aberration compensation design and control.

We are developing RF schemes to provide regeneration of the energy lost in the absorbers and to use synchrotron motion as an additional correction to chromatic effects [5], to include wedge engineering limitations, and to use realistic aberration correction magnets to provide guidance for simulations.

## APPENDIX A: HAMILTONIAN FRAMEWORK FOR PLANE BEAM BEND

Hamilton's method in dynamics provides a convenient and effective analytical technique of formulating, reducing and solving the equation of particle motion in an external field. A particular feature of this method is that it allows one to immediately recognize and utilize the dynamical invariants and canonical relationships between the invariants and some important characteristics of particle motion, such as orbital tunes, dispersions and others. The method is based on introduction of the Hamilton function of particle coordinates and canonical momenta in a chosen coordinate frame.

In classical dynamic theory, a conventional way of Hamilton's formulation is based on transformation from Lagrange-Euler equations to Hamilton's equations, either explicitly or through Hamilton-Jacobi method. Meanwhile, there exists a possibility to obtain the classical Hamilton function and canonical equations, starting with the Schrödinger or Klein-Gordon equation of quantum mechanics. In this way, the Hamilton formulation appears immediately as an initial or fundamental in classical mechanics (viewed as an asymptotic limit of the quantum description); the Lagrange-Euler formulation then is not inquired.

An ordinary Hamilton form is the energy function

$$H_t(\vec{P},\vec{r}) = \sqrt{p^2 + m^2} + A_t = \sqrt{(\vec{P} - \vec{A})^2 + m^2} + A_t,$$

with equations of motion

$$\frac{d}{dt}\vec{P} = -\frac{\partial}{\partial \vec{r}} H_t; \qquad \frac{d\vec{r}}{dt} = \frac{\partial}{\partial \vec{P}} H_t$$

### A. Hamilton function in s-representation of the reference orbit

In the case of a particle beam transported along a curved but plane reference orbit, it is convenient to consider beam path coordinate $s$ as a time argument, while the time can be treated as one of three independent coordinates $x, y, t$. As usual, we introduce *Frenet frame* with variables $x, y, s, t$:

$$\vec{r} = \vec{r}_0(s) + x\vec{e}_x + y\vec{e}_y; \qquad \vec{r}_0' = \vec{e}_s; \qquad \vec{e}_x' = K\vec{e}_s; \qquad \vec{e}_s' = -K\vec{e}_x$$

$$Kp_0 = B_0$$

The Hamiltonian function and equations of motion in this representation can be quickly derived using the covariant equation for the wave function $\Psi(\vec{r},t)$, or the relativistic Schrödinger equation

$$[(\hat{P}_t - A_t)^2 - (\vec{P} - \vec{A})^2 - m^2]\Psi(\vec{r},t) = 0$$

where $\hat{H}_t \equiv i\hbar \dfrac{\partial}{\partial t}$ and $\hat{\vec{P}} \equiv -i\hbar \dfrac{\partial}{\partial \vec{r}} \equiv \vec{\nabla}$

are the time and space components of four-vector momentum as a quantum operator. In Frenet frame differential operator $\vec{\partial}$ is written as

$$\vec{\nabla} = \vec{e}_x \frac{\partial}{\partial x} + \frac{\vec{e}_s}{1+Kx}\frac{\partial}{\partial s} + \vec{e}_y \frac{\partial}{\partial y}$$

In the quasi-classical limit, the Schrödinger equation can be rewritten, optionally, in two equivalent forms:

A)

$$i\hbar \frac{\partial}{\partial t}\Psi = [\sqrt{(\hat{\vec{P}} - \vec{A})^2 + m^2} + A_t]\Psi \equiv \hat{H}_t \Psi,$$

with corresponding classical equations of motion

$$\frac{d\hat{\vec{P}}}{dt} = -\frac{1}{i\hbar}[\hat{H}_t, \hat{\vec{P}}] \rightarrow \frac{d\vec{P}}{dt} = \{H_t, \vec{P}\} = -\frac{\partial}{\partial \vec{r}} H_t$$

$$\frac{d\vec{r}}{dt} = -\frac{1}{i\hbar}[\hat{H}_t, \vec{r}] \rightarrow \frac{d\vec{r}}{dt} = \{H_t, \vec{r}\} = \frac{\partial}{\partial \vec{P}} H_t$$

where $H_t$ is the conventional form of the Hamilton function (*t-representation*):

$$H_t \equiv \sqrt{(\hat{\vec{P}} - \vec{A})^2 + m^2} + A_t;$$

B)

$$i\hbar \frac{\partial}{\partial s}\Psi = -(1+Kx)[\sqrt{(\hat{H}_t - A_t)^2 - (\hat{\vec{P}}_\perp - \vec{A}_\perp)^2 - m^2} + A_s]\Psi \Rightarrow H_s \Psi,$$

with equations of motion (in a classical limit)

$$P'_x = -\frac{\partial}{\partial x}H_s; \quad x' = \frac{\partial}{\partial P_x}H_s; \quad P'_y = -\frac{\partial}{\partial y}H_s; \quad y' = \frac{\partial}{\partial P_y}H_s$$

$$H'_t = \frac{\partial}{\partial t}H_s; \qquad t' = \frac{\partial}{\partial s}H_s$$

where $H_s$ is the Hamilton function in *s-representation*:

$$H_s \equiv -(1+Kx)[\sqrt{(H_t - A_t)^2 - (\vec{P}_\perp - \vec{A}_\perp)^2 - m^2} + A_s]$$
$$= -(1+Kx)(\sqrt{p^2 - p_\perp^2} + A_s)$$

Thus, in s-representation the Hamiltonian coincides with the canonical momentum (with a reversed sign)

$$P_s = (1+Kx)(p_s + A_s)$$

taken as a function of energy and transverse momentum according to the covariant equation $E^2 - p^2 = m^2$:

$$H_s = -(1+Kx)(p_s + A_s) = -(1+Kx)(\sqrt{p^2 - \vec{p}_\perp^2} + A_s);$$

$$\vec{p}_\perp = \vec{P}_\perp - \vec{A}_\perp; \qquad p_x = P_x - A_x; \qquad p_y = P_y - A_y$$

$$H_t = E + A_t \qquad p^2 = E^2 - m^2$$

$$H_s = -(1+Kx)[\sqrt{(H_t - A_t)^2 - m^2 - (P_x - A_x)^2 - (P_y - A_y)^2} + A_s]$$

## B. Equations for vector potential of the magnetic field

We use the standard definition of vector potential of static magnetic field in vacuum:

$$\vec{B} = \vec{\nabla} \times \vec{A}$$

$$\vec{A} = A_s \vec{e}_s + A_x \vec{e}_x + A_y \vec{e}_y$$

$$\vec{\nabla}\vec{A} = 0, \vec{\nabla}^2 \vec{A} = 0$$

In Frenet frame we have the following equations:

$$\vec{\nabla}^2 \vec{A} = [\frac{1}{1+Kx}\frac{\partial}{\partial x}(1+Kx)\frac{\partial}{\partial x} + \frac{\partial^2}{\partial y^2} + \frac{1}{1+Kx}\frac{\partial}{\partial s}\frac{1}{1+Kx}\frac{\partial}{\partial s}]\vec{A} = 0$$

$$\frac{\partial A_s}{\partial s} + \frac{\partial}{\partial x}(1+Kx)A_x + (1+Kx)\frac{\partial}{\partial y}A_y = 0$$

Let us introduce the transverse part of the Laplace operator as

$$\Delta_\perp \equiv \frac{\partial^2}{\partial y^2} + \frac{1}{1+Kx}\frac{\partial}{\partial x}(1+Kx)\frac{\partial}{\partial x};$$

Then we rewrite equations for vector components as follows:

$$\Delta_\perp A_s + \frac{1}{(1+Kx)^2}[(1+Kx)\frac{\partial}{\partial s}\frac{1}{1+Kx}(\frac{\partial A_s}{\partial s} + KA_x) + K(\frac{\partial A_x}{\partial s} - KA_s)] = 0$$

$$\Delta_\perp A_x + \frac{1}{(1+Kx)^2}[(1+Kx)\frac{\partial}{\partial s}\frac{1}{1+Kx}(\frac{\partial A_x}{\partial s} - KA_s) - K(\frac{\partial A_s}{\partial s} + KA_x)] = 0$$

$$\Delta_\perp A_y + \frac{1}{1+Kx}\frac{\partial}{\partial s}\frac{1}{1+Kx}\frac{\partial}{\partial s}A_y = 0$$

## C. Expansion of the s-Hamiltonian

Below we will expand the vector potential and s-Hamiltonian in powers of deviations of particle momentum and transverse coordinates relatively the reference trajectory, in terms of the Frenet frame, including all terms up to the 4$^{th}$ power.

*Expansion of s-Hamiltonian on momentum vector*

$$H_s \approx -(1+Kx)\{p_0 + \Delta p - \frac{p_\perp^2}{2p_0}[1 - \frac{\Delta p}{p_0} + (\frac{\Delta p}{p_0})^2] - \frac{p_\perp^4}{8p_0^3} + A_s\}$$

$$\vec{p}_\perp = \vec{P}_\perp - \vec{A}_\perp$$

*Expansion of the vector potential of magnetic field*

$$A_s \approx A_{s1} + A_{s2} + A_{s3} + A_{s4};$$

$$\vec{A}_\perp \approx \vec{A}_{\perp 1} + \vec{A}_{\perp 2} + \vec{A}_{\perp 3}$$

*The first and second order terms of vector potential*

Longitudinal component:

$$A_{s1} = -B_0(s)x$$

$$A_{s2} = -\frac{1}{2}(\alpha x^2 + ny^2);$$

$$\Delta_\perp (A_{s1} + A_{s2})_0 \Rightarrow 0 \rightarrow \alpha = -n - K^2$$

$$A_{s2} = \frac{1}{2}[(K^2 + n)x^2 - ny^2]$$

Transverse components:

$$\vec{A}_{\perp 1} = 0$$

$$A_{x2} = 0; \qquad A_{y2} = \gamma_{y2} xy$$

$$\gamma_{y2} x - K'x = 0 \rightarrow \gamma_{y2} = K' \rightarrow A_{y2} = K'xy$$

*The third order terms of vector potential*

Longitudinal component:

$$\Delta_\perp (A_{s1} + A_{s2} + A_{s3}) = -\frac{\partial^2 A_{s1}}{\partial s^2} + K^2 A_{s1} = (K'' - K^3)x$$

$$(\frac{\partial^2}{\partial x^2} + \frac{\partial^2}{\partial y^2})A_{s3} + K\frac{\partial}{\partial x}A_{s2} - K^2 A_{s1} = (K'' - K^3)x$$

$$A_{s3} = \alpha_{s3}(x^3 - 3xy^2) + \hat{\alpha}_{s3}(x^3 + 3xy^2)$$

$$\hat{\alpha}_{s3} = \frac{1}{12}(K'' - 3K^3 - Kn)$$

Transverse components of the third order:

$$(\frac{\partial^2}{\partial x^2} + \frac{\partial^2}{\partial y^2})A_{x3} = \frac{\partial}{\partial s}KA_{s1} + K\frac{\partial}{\partial s}A_{s1}$$

$$\frac{\partial}{\partial x}A_{x3} + \frac{\partial}{\partial y}A_{y3} + Kx\frac{\partial}{\partial y}A_{y2} + \frac{\partial}{\partial s}A_{s2} = 0$$

$$(\frac{\partial^2}{\partial x^2} + \frac{\partial^2}{\partial y^2})A_{y3} + K\frac{\partial}{\partial x}A_{y2} = 0$$

$$A_{x3} = \alpha_{x3}(x^3 - 3xy^2) + \hat{\alpha}_{x3}(x^3 + 3xy^2)$$

$$A_{y3} = \hat{\beta}_{y3}(y^3 + 3x^2 y)$$

$$\alpha_{x3} = -\frac{1}{3}(KK' + \frac{1}{2}n')$$

$$A_{x3} = -\frac{1}{6}(2KK' + n')(x^3 - 3xy^2) - \frac{1}{4}KK'(x^3 + 3xy^2)$$

$$A_{y3} = K'xy - \frac{1}{12}KK'(y^3 + 3x^2 y)$$

*The fourth order terms of vector potential (longitudinal component only)*

$$(\frac{\partial^2}{\partial x^2} + \frac{\partial^2}{\partial y^2})A_{s4} = -K\frac{\partial A_{s3}}{\partial x} + K^2 x\frac{\partial A_{s2}}{\partial x} - K^3 x^2\frac{\partial A_{s1}}{\partial x} - (\frac{\partial^2}{\partial s^2} - K^2)A_{s2} +$$

$$2Kx(\frac{\partial^2}{\partial s^2} - K^2)A_{s1} + K'x\frac{\partial A_{s1}}{\partial s}$$

$$A_{s4} = \alpha_{s4}(x^4 + y^4 - 6x^2 y^2) + \beta_{s4}(x^4 - y^4) + \gamma_{s4}x^2 y^2$$

$$\gamma_{s4} = -\frac{3}{2}K\hat{\alpha}_{s3} + \frac{1}{4}a$$

$$\beta_{s4} = -\frac{1}{4}K\alpha_{s3} + \frac{1}{24}(K^2 n - n'') + \frac{1}{24}a$$

$$a \equiv \frac{9}{2}K^4 + K^2 n - 2K'^2 - 3KK''$$

*Expansion of the s-Hamiltonian on vector potential*

$$H_1 = -(Kp_0 - B_0)x - \Delta p \Rightarrow -\Delta p;$$

$$H_2 = -Kx\Delta p + \frac{P_\perp^2}{2p_0} - KxA_{s1} - A_{s2}$$

$$H_3 = \frac{P_\perp^2}{2p_0}(Kx - \frac{\Delta p}{p_0}) - KxA_{s2} - A_{s3} - \frac{\vec{P}_\perp \vec{A}_{\perp 2}}{p_0}$$

$$H_4 = -Kx(\frac{\vec{P}_\perp \vec{A}_{\perp 2}}{p_0} + \frac{P_\perp^2}{2p_0^2}\Delta p + A_{s3}) + \frac{\vec{A}_{\perp 2}^2}{2p_0} + \frac{P_\perp^4}{8p_0^3} - \frac{\vec{P}_\perp \vec{A}_{\perp 3}}{p_0} + \frac{P_\perp^2 (\Delta p)^2}{2p_0^3} - A_{s4}$$

**D. Second order Hamilton function for the linear optic design**

$$H_2 = -Kx\Delta p + \frac{P_x^2 + P_y^2}{2} + \frac{1}{2}[(K^2 - n)x^2 + ny^2]$$

Linear equations:

$$x'' + (K^2 - n)x = K\Delta p; \quad y'' + ny = 0$$

$$x = D\frac{\Delta p}{p} + x_b; \quad D'' + (K^2 - n)D = K$$

$$K(s) = K(s + \lambda_0); \quad n(s) = n(s + \lambda_0)$$

**E. The third power Hamiltonian terms**

Combining all the 3d power terms of the above expansions of the Hamiltonian and vector potential, we obtain the 3d power Hamiltonian part as follows:

$$H_3 = -\frac{1}{2}(p_x^2 + p_y^2)(q - Kx) - \frac{1}{2}K[(K^2 + n)x^3 - nxy^2]$$

$$-\alpha_3(x^3 + 3xy^2) - K'xyp_y - \frac{1}{3}n_{sext}(x^3 - 3xy^2)$$

$$\alpha_3 \equiv \frac{1}{12}(K'' - 3K^3 - Kn)$$

where the parameter $n_{sext}$ is the field index of the introduced sextupole components.

## F. The fourth power terms

Combining all the 4th power terms of the above expansions of the Hamiltonian and vector potential, we obtain the 4th power Hamiltonian part as follows:

$$H_4 = \frac{1}{8}(p_x^2 + p_y^2)^2 + \frac{1}{2}(K'xy)^2 - \alpha_3 K(x^4 + 3x^2 y^2) - \alpha_4 (x^4 - y^4)$$

$$- \beta_4 x^2 y^2 - \frac{1}{3} K n_{sext}(x^4 - 3x^2 y^2) - \frac{1}{4} n_{oct}(x^4 + y^4 - 6x^2 y^2)$$

Here we use the following notation:

$$\alpha_4 = \frac{1}{24}(K^2 n - n'' + \delta)$$

$$\beta_4 = -\frac{3}{2} K \alpha_3 + \frac{1}{4}\delta$$

$$\delta = \frac{9}{2} K^4 + K^2 n - 2K'^2 - 3KK'',$$

and $n_{oct}$ is the field index of the introduced octupole components.

More terms of the 4$^{th}$ power on transverse variables arrive as secondary effects from the 3d order Hamiltonian. They are generally formulated as resulting from a Poisson bracket as follows:

$$H_4^{(3)} = \frac{1}{2}\{H_3, \hat{H}_3\}$$

where

$$\hat{H}_3 \equiv \int H_3(s)\,ds,$$

with the Hamiltonian term $H_3$ considered as a function of time $s$ with constant particle trajectory parameters $a$ and $b$.

## APPENDIX B: TWIN-HELIX IMPLEMENTATION OF THE SNAKE CHANNEL

### A. Correlated Optics Condition

To ensure the beam's simultaneous focusing in both horizontal and vertical planes, the horizontal oscillation period $\lambda_x$ must be equal to or be a low-integer multiple of the vertical oscillation period $\lambda_y$. The PIC scheme also requires alternating dispersion D such that D is
- small at the beam focal points to minimize energy straggling in the absorber,
- non-zero at the absorber for emittance exchange to maintain constant longitudinal emittance,
- relatively large between the focal points to allow for aberration correction to keep the beam size small at the absorbers.

Given the above dispersion requirements, it is clear that $\lambda_x$ and $\lambda_y$ must also be low integer multiples of the dispersion period $\lambda_D$. Note that $\lambda_x$ and $\lambda_D$ should not be equal to avoid an unwanted resonance. Thus, the cooling channel optics must have correlated values of $\lambda_x$, $\lambda_y$ and $\lambda_D$:

$$\lambda_x = n\lambda_y = m\lambda_D \qquad\qquad 1.$$

e.g. $\lambda_x = 2\lambda_y = 4\lambda_D$ or $\lambda_x = 2\lambda_y = 2\lambda_D$.

## B. Orbital Dynamics

The PIC dynamics is very sensitive to magnetic fringe fields. One approach to finding a practical fringe-field-free solution is to use helical harmonics [2, 7]:

$$B_\varphi^n = \left(\frac{2}{nk}\right)^{n-1} \left.\frac{\partial^{n-1} B_\varphi^n}{\partial \rho^{n-1}}\right|_0 \cos(n[\varphi - kz + \varphi_0^n]) [I_{n-1}(nk\rho) - I_{n+1}(nk\rho)],$$

$$B_\rho^n = \left(\frac{2}{nk}\right)^{n-1} \left.\frac{\partial^{n-1} B_\varphi^n}{\partial \rho^{n-1}}\right|_0 \sin(n[\varphi - kz + \varphi_0^n]) [I_{n-1}(nk\rho) + I_{n+1}(nk\rho)],$$

$$B_z^n = -2\left(\frac{2}{nk}\right)^{n-1} \left.\frac{\partial^{n-1} B_\varphi^n}{\partial \rho^{n-1}}\right|_0 \cos(n[\varphi - kz + \varphi_0^n]) I_n(nk\rho),$$

where $B_\varphi$, $B_\rho$, and $B_z$ are the azimuthal, radial, and longitudinal helical magnetic field components, respectively, $n$ is the harmonic number (e.g. $n = 1$ is the dipole harmonic), and $k = 2\pi / \lambda_h$ is the helix wave number while $\lambda_h$ is the helix period.

One extensively-studied system based on helical field is the Helical Cooling Channel (HCC) [2]. However, the HCC is not suitable for PIC because it has constant dispersion magnitude. It was suggested [6] that alternating dispersion could be created by superimposing the HCC with an opposite-helicity helical dipole field with a commensurate characteristic period. However, with this approach, the periodic orbit solution is somewhat complicated and producing sufficiently large acceptance seems problematic.

Here we study a somewhat different configuration of magnetic fields, however, retaining the principle of Ref. [6] of creating an alternating dispersion by superimposing two helical-dipole fields with commensurate periods. We use a superposition of two equal-strength helical dipole harmonics with equal periods and opposite helicities ($k_1 = -k_2$) as a basis for our PIC channel design. Analogously to how combining two circularly-polarized waves produces a linearly-polarized one, the magnetic field in the mid-(horizontal)-plane of this configuration is transverse to the plane. This means that the periodic orbit is flat and lies in the mid-plane. The horizontal and vertical motions are uncoupled. This is a more conventional orbital dynamics problem than the one with a 3D reference orbit and coupled transverse motion.

Figure B1(a) shows an example of the periodic orbit solutions for 100 MeV/c $\mu^+$ and $\mu^-$ in a twin-helix channel with 1 m period and 0.741 T magnetic field strength of each helical dipole harmonic. The periodic orbit was determined numerically by locating the fixed point in the phase space. For this procedure, one only needed to consider the x-x' horizontal phase space and the procedure was further simplified by selecting a longitudinal position where x' was zero. Figure B1(b) shows the dispersion as a function of the longitudinal position for the $\mu^+$ solution shown in Fig. B1(a). Note that the dispersion has oscillatory behavior required for PIC. Note also that the dispersion period is equal to the helix period, i.e. $\lambda_D = \lambda_h$.

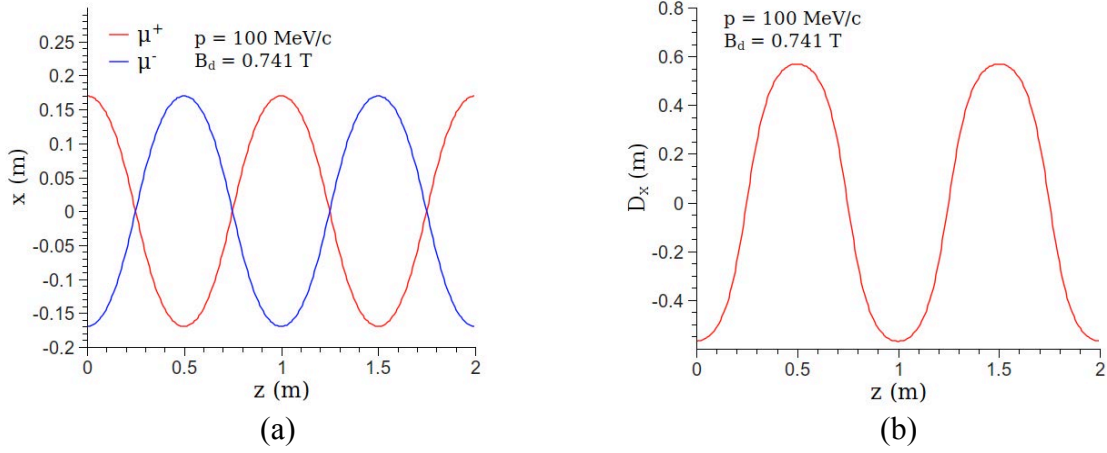

**FIGURE B1.** (a) Periodic orbits of 100 MeV/c $\mu^+$ and $\mu^-$ in a twin-helix channel with 1 m period and 0.741 T magnetic field strength of each helical dipole harmonic. (b) Dispersion behavior as a function of the longitudinal position for the $\mu^+$ solution in (a).

The transverse motion in a twin-helix consisting of two helical dipole harmonics only is stable around the periodic orbit in both dimensions as long as the helical dipole strength does not exceed a certain limiting value. Figures B2(a) and B2(b) show the periodic orbit amplitude and the betatron tunes, respectively, vs. the helical dipole strength $B_d$ for three different values of the helix period. In the calculations shown in Figs. B2(a) and B2(b), the strength $B_d$ was changed in small steps. On each step, the new periodic orbit was obtained as described above using the previous step's solution as the initial guess. The betatron tunes were extracted from a single-period linear transfer matrix, which was obtained numerically in terms of canonical coordinates. The canonical coordinates were calculated using an analytic expression for the helical magnetic field vector potential [7].

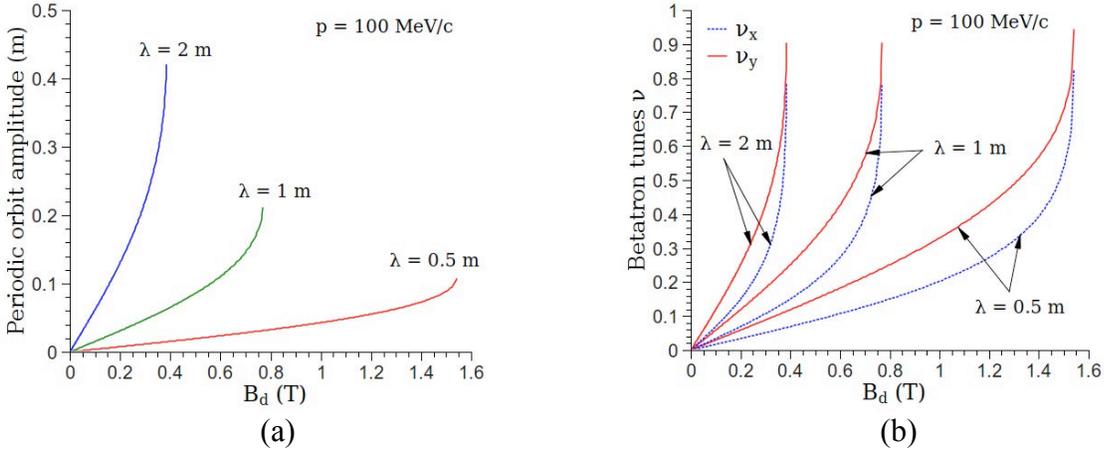

**FIGURE B2.** Periodic orbit amplitude (a) and horizontal and vertical betatron tunes (b) vs. the helical dipole strength.

Since the dispersion period is determined by the helix period $\lambda_D = \lambda_h$, the correlated optics condition of Eq. (i) imposes the following conditions on the betatron tunes: $\nu_x = \lambda_h/\lambda_x = \lambda_h/(m\lambda_D) = 1/m$ and $\nu_y = \lambda_h/\lambda_y = \lambda_h/(m\lambda_D/n) = n/m$, e.g. $\nu_x = 0.5$, $\nu_y = 1$ or $\nu_x = 0.25$, $\nu_y = 0.5$. Examining Fig. 3(b) shows that it is not possible to satisfy these conditions by adjusting $\lambda_h$ and $B_d$. Thus, we introduced a straight normal quadrupole to redistribute focusing between the horizontal and vertical dimensions. One subtlety is that, in addition to changing the focusing properties of the lattice, the quadrupole also changes the periodic reference orbit. The helical dipole strength $B_d$ and the quadrupole gradient $\partial B_y/\partial x$ were iteratively adjusted until, at $B_d = 1.303$ T and $\partial B_y/\partial x = 1.153$ T/m, we achieved the correlated optics condition with $\nu_x = 0.25$ and $\nu_y = 0.5$. Having $\nu_x = 0.5$ and $\nu_y = 1$ would be more beneficial by allowing shorter spacing between the absorbers but it was not possible to adjust these tune values because of a strong parametric resonance at $\nu_y = 1$. We also attempted tuning the correlated optics condition by using a helical quadrupole pair instead of a straight quadrupole but that configuration did not seem compatible with the correlated optics requirements. Figures 4(a) and 4(b) show the dependence of the betatron tunes on $B_d$ at $\partial B_y/\partial x = 1.153$ T/m and on $\partial B_y/\partial x$ at $B_d = 1.303$ T, respectively. The crossing lines indicate the correlated optics condition. Note that the straight quadrupole introduces an asymmetry into the magnetic field so that the correlated optics condition is satisfied for one muon charge only.

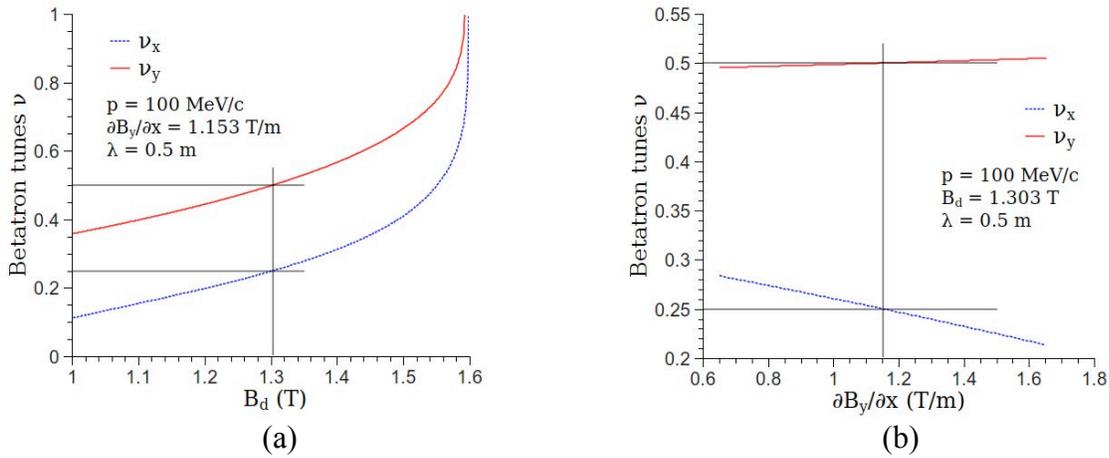

**FIGURE B3.** Horizontal and vertical betatron tunes vs. the helical dipole strength $B_d$ at $\partial B_y/\partial x = 1.153$ T/m (a) and the straight quadrupole gradient $\partial B_y/\partial x$ at $B_d = 1.303$ T (b).

We next studied the dependence of the periodic orbit and of the betatron tunes on the muon momentum as shown in Figs. B4(a) and B5(b). These studies demonstrated large momentum acceptance of the twin-helix channel. We found that, with correlated optics, the dispersion amplitude $D_{x\,max} = p\,\partial x_{max}/\partial p$ was 0.098 m and the horizontal and vertical chromaticities were $\xi_x = p\,\partial \nu_x/\partial p = -0.646$ and $\xi_y = -0.798$, respectively. The correlated optics parameters scale with the muon momentum and helix period in the following way:

$$B_d \propto p/\lambda; \quad \partial B_y/\partial x \propto p/\lambda^2; \quad x_{max}, D_x \propto \lambda; \quad \xi_x, \xi_y \propto \text{const.} \qquad 2.$$

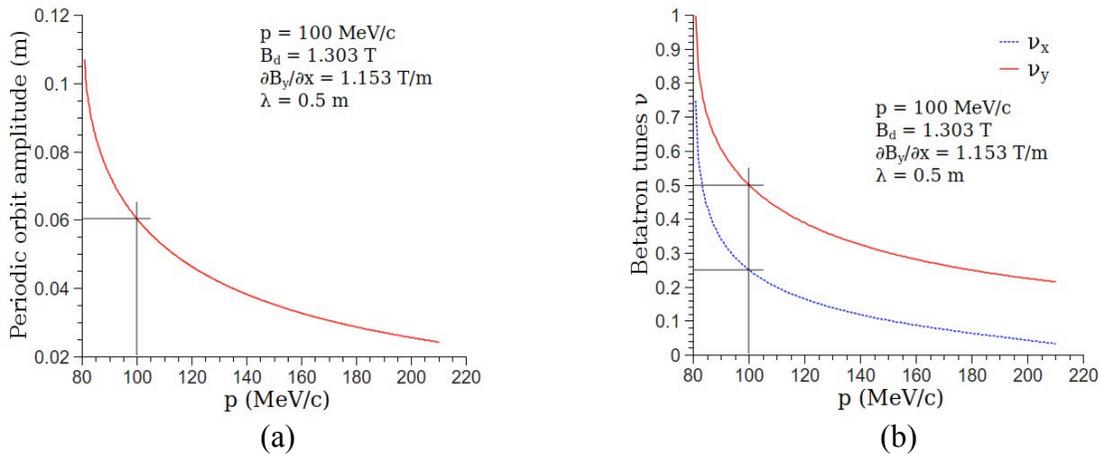

**FIGURE B4.** Periodic orbit amplitude (a) and betatron tunes (b) vs. the muon momentum with correlated optics.

## C. G4beamline Simulation

To estimate the dynamic aperture of the correlated optics channel, we tracked $10^5$ 100 MeV/c muons through 100 periods of the channel using the GEANT-based G4beamline program [8]. The initial muon beam was monochromatic, parallel and uniformly distributed within a 10 cm × 10 cm square. Figures B5(a) and B5(b) show the initial and final transverse muon distributions, respectively. The initial positions of the particles that were not lost after 100 periods are shown in Fig. B5(a) in blue. Figures B5(a) and B5(b) suggest a very large dynamic aperture and therefore high transmission efficiency of the twin-helix channel.

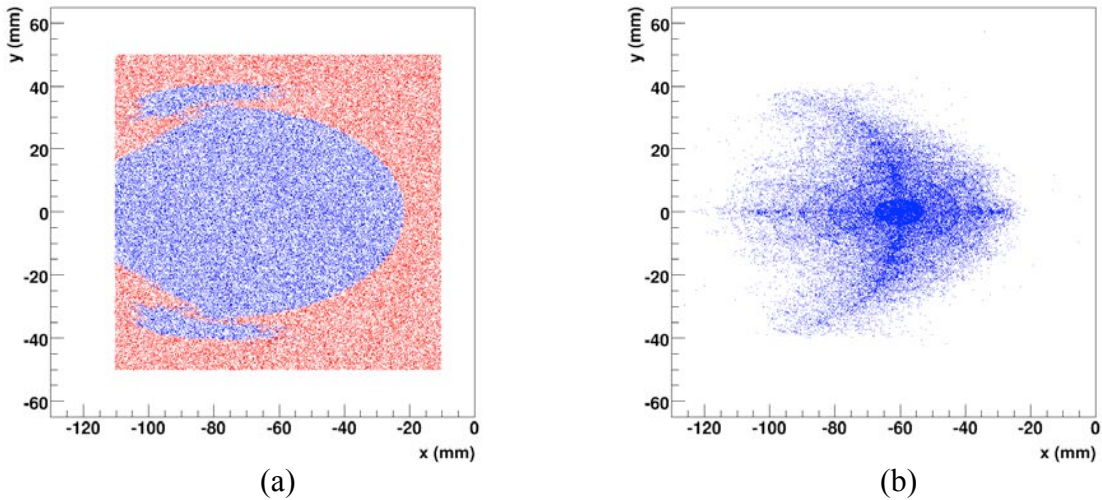

**FIGURE B5.** Initial (a) and final (b) transverse muon beam distributions in a G4beamline tracking study.

## D. Possible Practical Implementation

In practice, the required magnetic field configuration can be obtained by winding two separate coaxial layers of helical conductors and coaxially superimposing a straight quadrupole as shown in Fig. B6(a). The helical conductors constituting the two layers have the same special periods and opposite helicities. Within each layer, the currents in the helical conductors vary azimuthally as $\cos(\phi)$. Note that the two layers do not have to have the same radius. The difference in the radii can be accounted for by adjusting the layers' currents. Another approach to producing the desired magnetic field is to combine two coaxial layers consisting of series of tilted current loops and overlay a coaxial straight quadrupole as shown in Fig. B6(b). The inclined loops comprising the two different layers are tilted in opposite directions. The current in the loops of each layer varies longitudinally as cos($kz$). Such a technology, without longitudinal current variation, is used for making constant-field dipoles.

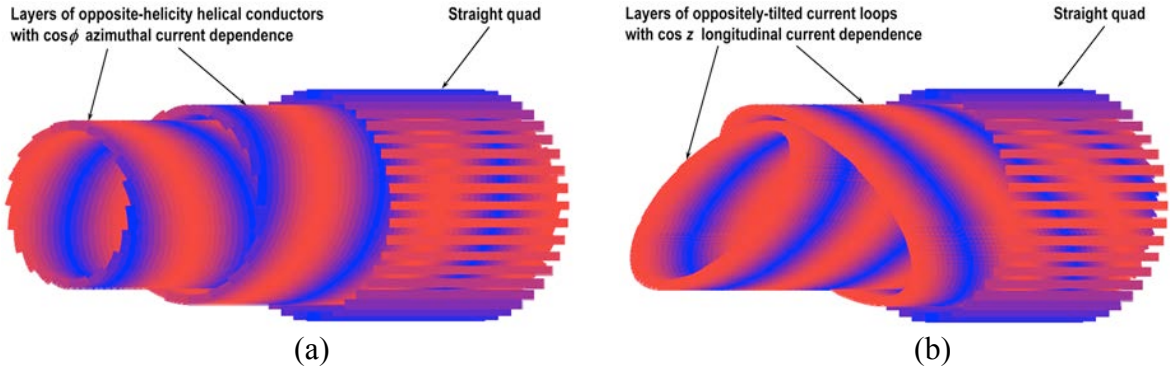

(a) (b)

**FIGURE B6.** Conceptual diagrams of possible practical implementations of the twin-helix channel (a) using a combination of two helical conductor layers and a straight quadrupole and (b) using a combination of two layers of tilted current loops and a straight quadrupole. The colors represent current variation in the conductiors.

## APPENDIX C: STUDIES OF THE TWIN HELIX PARAMETRIC RESONANCE IONIZATION COOLING CHANNEL WITH COSY INFINITY

The twin helix channel is simulated using COSY Infinity, a DA-based code that allows for calculation of non-linear effects to arbitrary order [7]. This paper details a linear simulation of this channel, with and without stochastic effects, and studies cooling efficiency with and without the effects of PIC. The linear simulation provides a baseline for ideal cooling in the channel if nonlinear aberrations in the channel have been fully corrected.

### A. Simulation Parameters

Table C1 details the parameters used in this linear simulation. The basic cell consists of a continuous straight quadrupole field superimposed upon a pair of helical harmonic dipole fields to establish the correlated optics condition. Wedge absorbers, made of beryllium with a central thickness of 2 cm and a gradient of 30%, are placed every 4 meters in the channel at a location of small but non-zero dispersion. Idealized RF cavities are placed 3 cm after the center of each wedge. COSY INFINITY calculates the transfer map for a 4-meter long cell (from the center of a wedge absorber through the center of the next wedge absorber). Figure 1 illustrates the geometry of this cell.

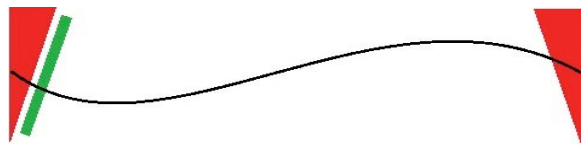

**Figure C1.** Schematic of Single Twin Helix Cell.

The total transfer map for the twin helix channel is obtained by composing the maps for each of the cells that make up the channel upon one another. The orbit of the reference particle (250 MeV/c muon) is periodic from the beginning to the end of each cell.

Table C1: 4-meter Cell Parameters

| | |
|---|---|
| H. Dipole Field | 1.63 T |
| H. Dipole wavelength | 1 meter |
| Continuous Quadrupole Field | .72 T/m |
| H. Quadrupole Field (Horizontal Lenses) | .02 T/m |
| H. Quadrupole wavelength | 2 meters |
| H. Quadrupole Field (Vertical Lenses) | .04 T/m |
| H. Quadrupole wavelength | 1 meter |
| RF Voltage | -12.5 MeV |
| RF Frequency | 201.25 MHz |
| RF Phase | 30 Degrees |

**B. Simulation of the Parametric Resonance Condition in the Twin Helix Channel**

To induce the resonance condition for PIC in the twin helix channel, two independent pairs of helical harmonic quadrupole fields (parametric lenses) are used; one pair induces resonance in the horizontal plane, the other in the vertical plane. The resonances induced by these fields create a hyperbolic fixed point; i.e., motion of particles relative to the reference orbit at the center of the wedge absorber becomes hyperbolic rather than elliptical. Figures C2a-b show this condition in the basic cell (without wedge absorber or RF) when a test particle that is offset both horizontally and vertically in both position and angle relative to the reference orbit by 2 cm and 130 mrads respectively. With the parametric lenses, the position offset is quickly minimized at the expense of a rapid blowup in the angle offset.

**C. Simulation of Ionization Cooling in the Twin Helix Channel**

The next stage in the simulations adds wedge absorbers and RF every 4 meters. Figures C3a-b show simulations with and without the parametric lenses to demonstrate the effects of ionization cooling with and without the PIC condition. The simulations demonstrate the expected results. With only ionization cooling, initial cooling effects are primarily to angle, and only later to the positional offset. With PIC the cooling effects are primarily to position offset, i.e., spot size of the beam. The increase in angle offset is minimized thru the ionization cooling effects of the wedge absorber.

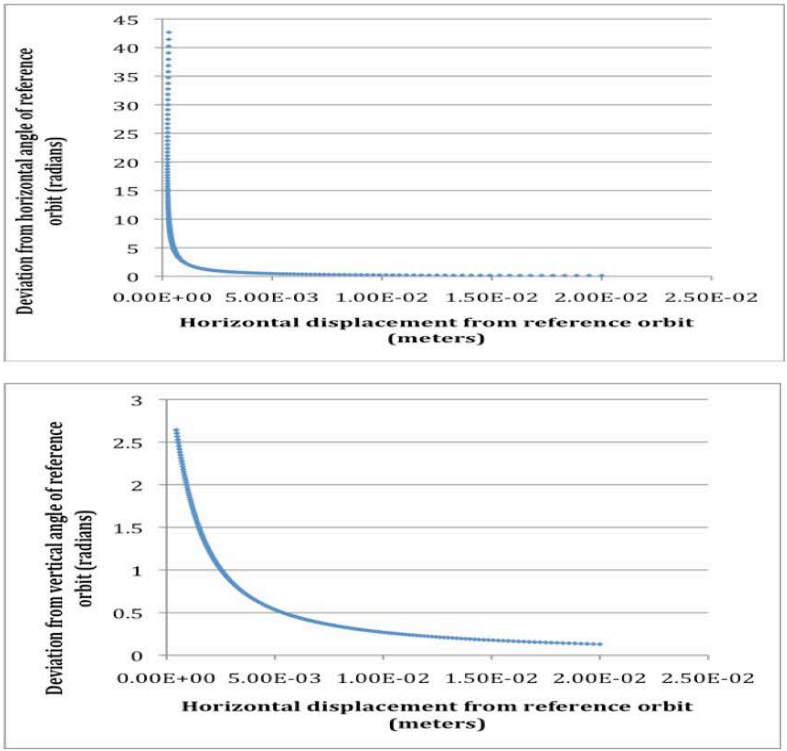

**Figure C2a-b.** Single particle launched with a horizontal and vertical offsets of 2 cm and .130 radians from the reference orbit and tracked for 200 cells.

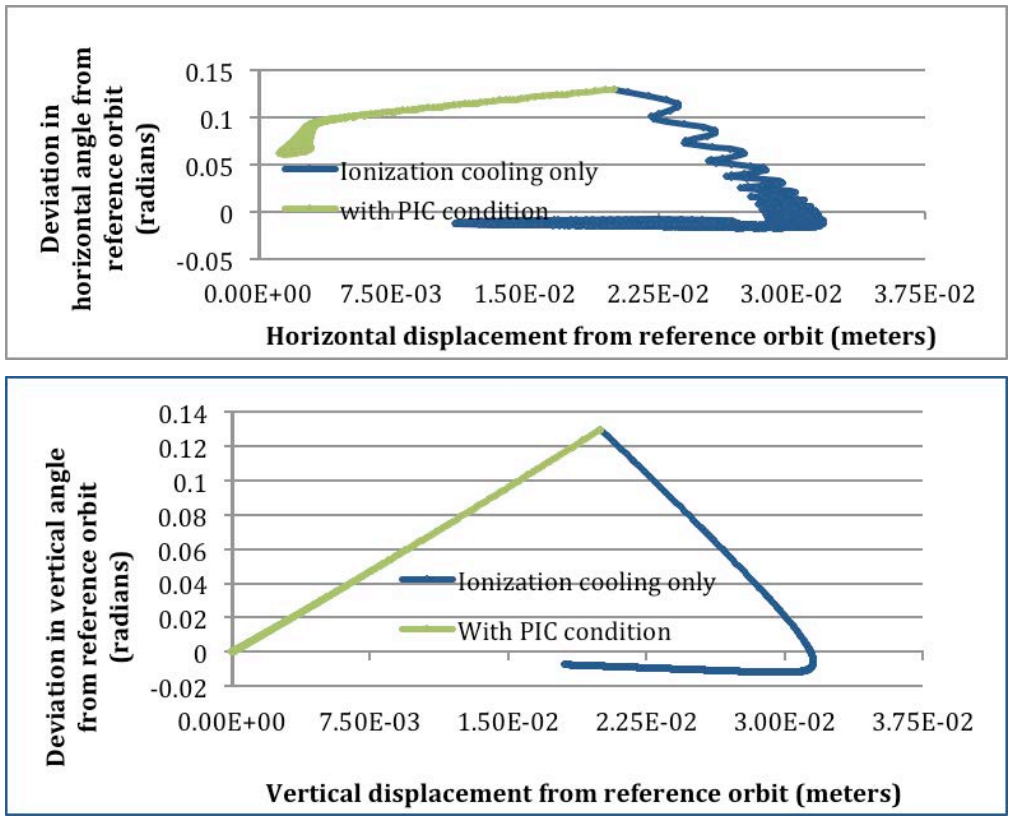

**Figures C3a-b.** Single particle tracked through 1000 cells with PIC condition and with ionization cooling only.

### D. Simulation of Stochastic Effects in the Twin Helix Channel

Stochastic effects of multiple scattering and energy straggling within the wedge absorber were then added to the simulations. Figures C4a-b show the results of combining ionization cooling with stochastic effects on a single particle initially offset both horizontally and vertically in both position and angle relative to the reference orbit by 2 cm and 130 mrad respectively.

As expected, the test particle is cooled until equilibrium is reached when cooling has been balanced with the effects of multiple scattering and energy straggling [6].

A distribution of test particles was also used to test cooling effects in the full simulation of the linear channel. The initial distribution uses a sigma of 2 cm in positions, 130 mrad in angles, and 1% spread in energy from the reference particle. The distribution is also spread over a bunch length of ± 3 cms relative to the reference particle. Figure C5 shows the 2D emittance change in the system calculated from the distribution. The horizontal and vertical 2D emittances are both reduced until equilibrium is reached. Longitudinal emittance is determined from deviation in path length and energy from the reference particle. Once the transverse emittance has reached equilibrium, the increases in longitudinal emittance contribute to heating in the beam distribution. Total 6D emittance for the distribution is plotted in Figure 6 with and without parametric lenses to induce the PIC condition. Figure C7 shows the cooling factor for the channel with and without the PIC condition.

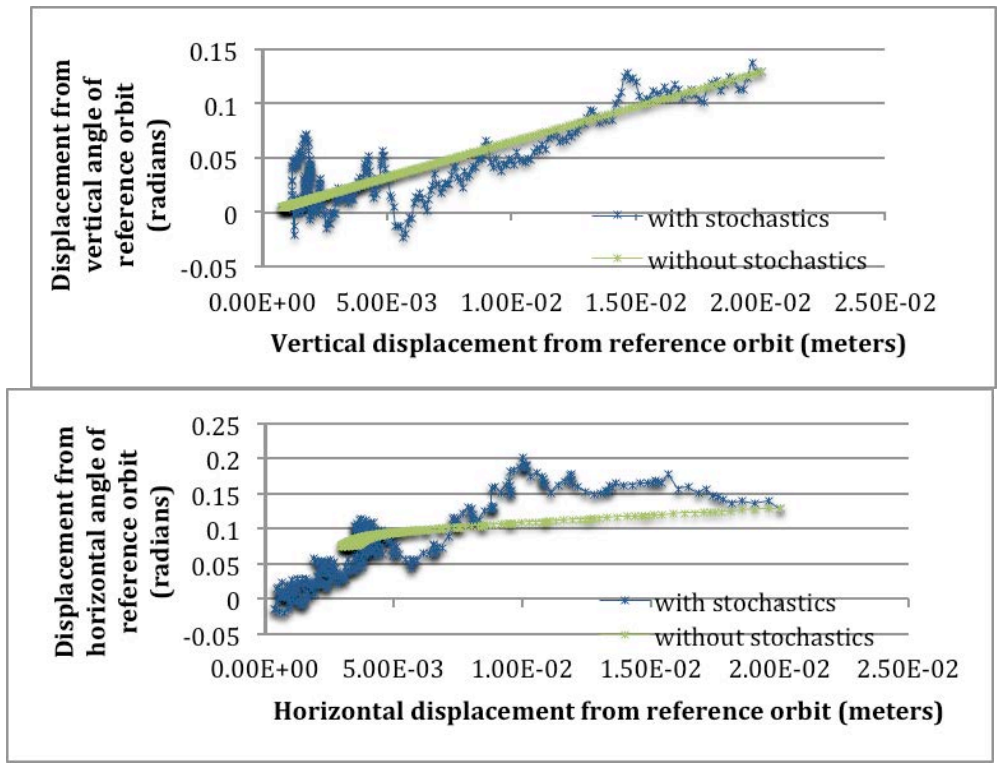

**Figure C4a-b.** Single particle tracked showing cooling thru 350 cells with and without stochastic effects.

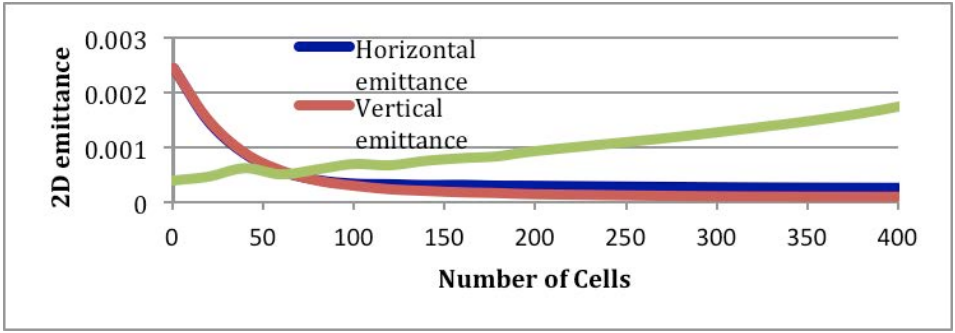

**Figure C5**. Emittance reduction for a distribution of 1000 particles tracked through the twin-helix channel with the PIC condition and stochastic effects.

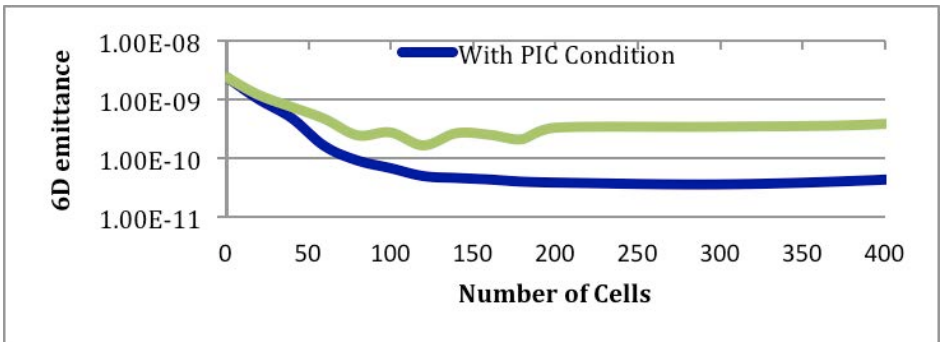

**Figure C6.** Comparison of 6D emittance reduction with and without the PIC condition.

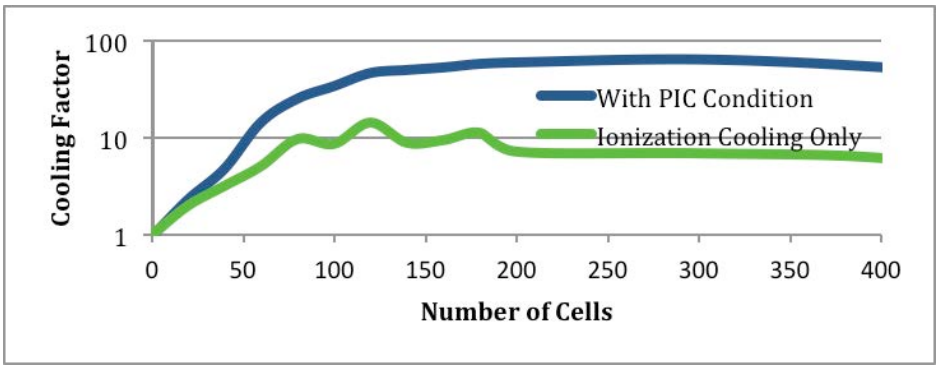

**Figure C7.** Comparison of cooling factor (ratio of initial to final 6D emittance) with and without the PIC condition.

The determinant of the transfer map for the cell, a 6x6 matrix in the linear case, can also be used to show transverse and 6D cooling in the system [10]. The determinant for the transfer matrix for this test channel is 0.945372. The determinant of the transverse 4x4 quadrant of the transfer matrix is 0.986054.

**E. Conclusions and Future Work**

Current simulations in COSY INFINITY have demonstrated that the linear model with stochastic effects of the twin helix channel achieves the resonance condition for PIC, as well as 6D cooling. Future simulations will determine the optimal parameters for this linear model, including cell length, magnet strengths, helicity and phase shifts for the helical harmonic magnets. Wedge gradients and thickness, as well as RF placement and parameters, will also be optimized. Next, studies will determine the largest nonlinear aberrations affecting this optimized twin-helix channel and the dependence these aberrations have on higher order helical harmonic and continuous multipole fields of differing strength, helicity and phase. The cooling efficiency of a system with corrected higher order effects can then be measured and compared with competing 6D cooling methods.